\documentclass[english]{revtex4-1}
\usepackage[T1]{fontenc}
\usepackage[utf8]{inputenc}
\usepackage[a4paper]{geometry}
\usepackage{babel}
\usepackage{float}
\usepackage{amsmath}
\usepackage{amssymb}
\usepackage{graphicx}

\makeatletter
\usepackage{bbm}
\DeclareMathOperator{\erf}{erf}

\makeatother

\begin{document}

\title{An analytic approximation of the feasible space of metabolic networks}

\author{Alfredo Braunstein}
\email{alfredo.braunstein@polito.it}

\affiliation{DISAT, Politecnico di Torino, 10129 Torino, Italy}

\affiliation{Human Genetics Foundation-Torino, 10126 Torino, Italy}

\affiliation{Collegio Carlo Alberto, 10024 Moncalieri, Italy}

\author{Anna Paola Muntoni}

\affiliation{DISAT, Politecnico di Torino, 10129 Torino, Italy}

\author{Andrea Pagnani}

\affiliation{DISAT, Politecnico di Torino, 10129 Torino, Italy}

\affiliation{Human Genetics Foundation-Torino, 10126 Torino, Italy}

\affiliation{Istituto Nazionale di Fisica Nucleare (INFN) Via Pietro Giuria, 10125, Torino, Italy}

\begin{abstract}
Assuming a steady-state condition within a cell, metabolic fluxes
satisfy an under-determined linear system of stoichiometric\emph{
}equations. Characterizing the space of fluxes that satisfy such equations
along with given bounds (and possibly additional relevant constraints)
is considered of utmost importance for the understanding of cellular
metabolism. Extreme values for each individual flux can be computed
with Linear Programming (as Flux Balance Analysis), and their marginal
distributions can be approximately computed with Monte-Carlo sampling.
Here we present an approximate analytic method for the latter task
based on Expectation Propagation equations that does not involve sampling
and can achieve much better predictions than other existing analytic
methods. The method is iterative, and its computation time is dominated
by one matrix inversion per iteration. With respect to sampling, we
show through extensive simulation that it has some advantages including
computation time, and the ability to efficiently fix empirically estimated
distributions of fluxes. 
\end{abstract}
\maketitle

\section*{Introduction}

The metabolism of a cell entails a complex network of chemical reactions
performed by thousands of enzymes that continuously process intake
nutrients to allow for growth, replication, defense and other cellular
tasks \cite{nelson2008lehninger}. Thanks to the new high-throughput
techniques and comprehensive databases of chemical reactions, large
scale reconstructions of organism-wide metabolic networks are nowadays
available. Such reconstructions are believed to be accurate from a
topological and stoichiometric viewpoint (e.g. the set of metabolites
targeted by each enzyme, and their stoichiometric ratio). For the
determination of reaction rates, large-scale constraint based approaches
have been proposed \cite{palsson2015systems}. Typically, such methods
assume a steady-state regime in the system where metabolite concentrations
remain constant over time (mass-balance condition). A second type
of constraints limit the reaction velocities and their direction.
In full generality the topology of a metabolic network is described
in terms of the chemical relations between the $M$ metabolites and
$N$ reactions. In mathematical terms we can define a $M\times N$
stoichiometric matrix $\mathbf{S}$ in which rows correspond to the
stoichiometric coefficients of the corresponding metabolites in all
reactions. A positive (resp. negative) \textbf{$S_{ij}$} term indicates
that metabolite $i$ is created (resp. consumed) by reaction $j$.
Assuming mass-balance and limited interval of variation for the different
reactions, we can cast the problem in terms of finding the set of
fluxes $\boldsymbol{\nu}\in\mathbb{R}^{N}$ compatible with the following
linear system of constraints and inequalities:

\begin{eqnarray}
\mathbf{S}\boldsymbol{\nu} & = & \boldsymbol{b}\label{eq:soichio}\\
\boldsymbol{\nu}^{inf}\leq & \boldsymbol{\nu} & \boldsymbol{\leq\nu}^{sup}\label{eq:bounds}
\end{eqnarray}
where $\boldsymbol{b}\in\mathbb{R}^{M}$ is the known set of intakes/uptakes,
and the pair $\boldsymbol{\nu}{}^{inf},\boldsymbol{\nu}{}^{sup}$
represent the extremes of variability for the variables of our problem.
Only in few cases the extremes are experimentally accessible, in the
remaining ones they are fixed to arbitrarily large values. It turns
out that $N\geq M$, and the system is typically under-determined.
As an example, the RECON1 model of Homo Sapiens has $N=2469$ fluxes
(\emph{i.e. }variables) and $M=1587$ metabolites (\emph{i.e. }equations).
The mass-balance constraints and the flux inequalities encoded in
Eqs.~\eqref{eq:soichio}-\eqref{eq:bounds} define a convex bounded
polytope, which constitutes the space of all feasible solutions of
our metabolic system. 

The most widely used technique to analyze fluxes in large scale metabolic
reconstruction is Flux Balance Analysis (FBA) \cite{varma_metabolic_1994,kauffman_advances_2003}
where a linear objective function, typically the biomass or some biological
proxy of it is introduced, and the problem reduces to find the subspace
of the polytope which optimizes the objective function. If this subspace
consists in only one point, the problem can be efficiently solved
using linear programming. FBA has been successfully applied in many
metabolic models to predict specific phenotypes under specific growth
condition (e.g. bacteria in the exponential growth phase). However,
if one is interested in describing more general growth conditions,
or is interested in other fluxes than the biomass \cite{Suthers2007},
different computational strategies must be envisaged \cite{Marchiori_2014,fernandez-de-cossio-diaz_fast_2016,DeMartinoParisi2014}.

As long as no prior knowledge is considered, each point of the polytope
is an equally viable metabolic phenotype of the biological system
under investigation. Therefore, being able to sample high-dimensional
polytopes becomes a theoretical problem with concrete practical applications.
From a theoretical standpoint the problem is known to be \#P-hard
\cite{Dyer_Frieze1988} and thus an approximate solution to the problem
must be sought. A first class of Monte Carlo Markov chain sampling
techniques available to analyze large-dimensional polytopes was originally
proposed three decades ago \cite{Smith1984} and falls under the name
of Hit-and-Run (HR) \cite{turchin_computation_1971}. Basically, it
consists on iteratively collecting samples by choosing random directions
from a starting point belonging to the polytope. Unfortunately polytopes
defined by large-scale metabolic reconstructions are typically ill-conditioned
(i.e. some direction of the space are far more elongated than others),
and improved HR techniques to overcome this problem have been proposed
\cite{kaufman1998} and implemented in the context of metabolic modeling
\cite{Schellenberger2011,Marchiori_2014,DeMartinoParisi2014}. Despite
the fact that these dynamic sampling strategies are often referred
as uniform random samplers, the uniformity of the sampling is guaranteed
only in an asymptotic sense, and often establishing in practice how
long a simulation should be run and how frequently the measurement
should be taken for a given instance of the problem requires extensive
preliminary simulations which make their use very difficult under
general conditions. Note also that the problem of assessing how perturbations
of network parameters affect the structure of the polytope is often
of practical importance; e.g. changing extremal flux values for studying
growth rate curves or enzymopaties \cite{Price2004}. In these situations
in principle the convergence time of the algorithm should be established
independently for each new value of the parameter. Another limitation
of this class of sampling strategies is the difficulty of imposing
other constraints \cite{schellenberger_use_2009} such as the experimentally
measured distribution profiles of specific subset of fluxes (typically
biomass and/or in-take/out-take of the network), a particularly timely
issue given the recent breakthrough of metabolic measurements in single
cell \cite{TaheriAraghi2015}, although recent attempts in this direction\textbf{
}exist \cite{de_martino_scalable_2012,de_martino_growth_2016}.

Recently, alternative statistical methods based on message passing
(MP) techniques (also known as cavity or Bethe approximation in the
context of statistical mechanics) \cite{Mezard_Montanari_book} have
been proposed \cite{braunstein_estimating_2008,braunstein_space_2008,massucci2013,fernandez-de-cossio-diaz_fast_2016,font-clos_weighted_2012},
allowing for sampling of the polytope orders of magnitude faster than
HR methods, under two main conditions: (i) the graphical structure
of the graph must be a tree or, at least, locally tree-like (i.e.
without short loops), (ii) the rows of the stoichiometric matrix $\mathbf{S}$
should be statistically uncorrelated. Unfortunately, neither assumption
is really fulfilled by large-scale metabolic reconstructions. To give
an example, consider the rows of the stoichiometric matrix for \textit{ecoli-core}
model \cite{orth_reconstruction_2010}. The rows corresponding to
the adenosine-diphosphate (ADP) and adenosine-triphosphate (ATP) appear
strongly correlated as both metabolites commonly appear in 11 reactions;
the same apply for the intracellular water and hydrogen ion that have
10 reactions in common. For these reasons MP methods suffer from all
kind of convergence and accuracy problems.

In this work we propose a new Bayesian inference strategy to analyze
with unprecedented efficiency large dimensional polytopes. The use
of a Bayesian framework allows us to map the original problem of sampling
the feasible space of solutions of Eqs.~\eqref{eq:soichio}-\eqref{eq:bounds}
into the inference problem of the joint distribution of metabolic
fluxes. Linear and inequality constraints will be encoded within the
likelihood and the prior probabilities that via Bayes theorem provide
a model for the posterior $P\left(\boldsymbol{\nu}|\boldsymbol{b}\right)$.
The goal of this work is to determine a tractable multivariate probability
density $Q\left(\boldsymbol{\nu}|\boldsymbol{b}\right)$ able to accurately
approximate the posterior even in the case of strongly row-correlated
stoichiometric matrices. This strategy relies on an iterative and
local refinement of the parameters of $Q\left(\boldsymbol{\nu}|\boldsymbol{b}\right)$
that falls into the class of Expectation Propagation (EP) algorithms.
We report results of EP for representative state-of-the-art models
of metabolic networks in comparison with HR estimate, showing that
EP can be used to compute marginals in a fraction of the computing
time needed by HR. We also show how the technique can be efficiently
adapted to incorporate the estimated growth rate of a population of
\textsl{Escherichia Coli}.

\section*{Results}

\subsection*{Formulation of the problem}

We are going to formulate an iterative strategy to solve the problem
of finding a multivariate probability measure over the set of fluxes
$\boldsymbol{\nu}$ compatible with Eqs.~\eqref{eq:soichio}-\eqref{eq:bounds}.
For a vector of fluxes satisfying bounds \ref{eq:bounds}, we can
define a quadratic energy function $E\left(\mathbf{\boldsymbol{\nu}}\right)$
whose minimum(s) lies on the assignment of variables $\boldsymbol{\nu}$
satisfying the stoichiometric constraints in Eq. \eqref{eq:soichio}:
\begin{equation}
E\left(\boldsymbol{\nu}\right)=\frac{1}{2}\left(\mathbf{S}\boldsymbol{\nu}-\boldsymbol{b}\right)^{T}\left(\mathbf{S}\boldsymbol{\nu}-\boldsymbol{b}\right)\label{eq:energy}
\end{equation}

We define the likelihood of observing $\boldsymbol{b}$ given a set
of fluxes $\boldsymbol{\nu}$ as a Boltzmann distribution:

\begin{equation}
P\left(\boldsymbol{b}|\boldsymbol{\nu}\right)=\left(\frac{\beta}{2\pi}\right)^{\frac{M}{2}}e^{-\frac{\beta}{2}\left(\mathbf{S}\boldsymbol{\nu}-\boldsymbol{b}\right)^{T}\left(\mathbf{S}\boldsymbol{\nu}-\boldsymbol{b}\right)}\label{eq:likelihood}
\end{equation}
where $\beta$ is a positive parameter, the ``inverse temperature''
in statistical physics jargon, that governs the penalty of whose configurations
of fluxes that are far from the minimum of the energy. In a Bayesian
perspective one can consider the posterior probability of observing
$P\left(\boldsymbol{\nu}|\boldsymbol{b}\right)$ as:

\begin{equation}
P\left(\boldsymbol{\nu}|\boldsymbol{b}\right)=\frac{P\left(\boldsymbol{b}|\boldsymbol{\nu}\right)P\left(\boldsymbol{\nu}\right)}{P\left(\boldsymbol{b}\right)}\label{eq:bayes}
\end{equation}
 where the prior
\begin{equation}
P\left(\boldsymbol{\nu}\right)=\prod_{n=1}^{N}\psi_{n}\left(\nu_{n}\right)=\prod_{n=1}^{N}\frac{\mathbb{I}\left(\nu_{n}\in\left[\nu_{n}^{inf},\nu_{n}^{sup}\right]\right)}{\nu_{n}^{sup}-\nu_{n}^{inf}}\label{eq:prior}
\end{equation}
enforces the bounds over the allowed range of fluxes. The function
$\mathbb{I}\left(\nu_{n}\in\left[\nu_{n}^{inf},\nu_{n}^{sup}\right]\right)$
is an indicator function that takes value $1$ if $\nu_{n}\in\left[\nu_{n}^{inf},\nu_{n}^{sup}\right]$
and $0$ otherwise. We finally obtain the following relation for the
posterior:

\begin{equation}
P\left(\boldsymbol{\nu}|\boldsymbol{b}\right)=\frac{1}{P\left(\boldsymbol{b}\right)}\left(\frac{\beta}{2\pi}\right)^{\frac{M}{2}}e^{-\frac{\beta}{2}\left(\mathbf{S}\boldsymbol{\nu}-\boldsymbol{b}\right)^{T}\left(\mathbf{S}\boldsymbol{\nu}-\boldsymbol{b}\right)}\prod_{n=1}^{N}\psi_{n}\left(\nu_{n}\right)\label{eq:posterior}
\end{equation}
and eventually we will investigate the $\beta\rightarrow\infty$ limit.
Neglecting terms that do not depend on $\boldsymbol{\nu}$, the posterior
takes the form of

\begin{equation}
P\left(\boldsymbol{\nu}|\boldsymbol{b}\right)\propto e^{-\frac{\beta}{2}\left(\mathbf{S}\boldsymbol{\nu}-\boldsymbol{b}\right)^{T}\left(\mathbf{S}\boldsymbol{\nu}-\boldsymbol{b}\right)}\prod_{n=1}^{N}\psi_{n}\left(\nu_{n}\right)
\end{equation}
where we have not explicitly reported the normalization constant.
By marginalization of Eq. \eqref{eq:posterior} one can determine
the marginal posterior $P_{n}\left(\nu_{n}|\boldsymbol{b}\right)$
for each flux $n\in\left\{ 1,\ldots,N\right\} $. However, performing
this computation naively would require the calculation of a multiple
integral that is in principle computationally very expensive and cannot
be performed analytically in an efficient way.

A standard way of approximately computing $P\left(\boldsymbol{\nu}|\boldsymbol{b}\right)$
is through sampling methods, such as the HR technique. The accuracy
obtained with HR depends of course on the number of samples, and sampling
accurately can be very time consuming. In the following we develop
an analytic approach to approximately compute marginal posteriors
which is able to achieve results as accurate as the HR sampling technique
for a large number of sampled points in a fraction of the computing
time. But first, we will describe as a warm-up a naive analytic method
to approximately compute marginal distributions $P_{n}\left(\nu_{n}|\boldsymbol{b}\right)$.

\subsection*{A non-adaptive approach\label{sec:Local}}

As a first approximation one can think of replacing each exact prior
$\psi_{n}\left(\nu_{n}\right)$ with a single Gaussian distribution
$\mathcal{\phi}_{n}\left(\nu_{n};a_{n},d_{n}\right)=\frac{e^{-\frac{\left(\nu_{n}-a_{n}\right)^{2}}{2d_{n}}}}{\sqrt{2\pi d_{n}}}$
whose statistics, i.e. the mean and the variance, are constrained
to be equal to the one of $\psi_{n}\left(\nu_{n}\right)$. That is

\begin{equation}
\begin{cases}
a_{n} & =\left\langle \nu_{n}\right\rangle _{\psi_{n}\left(\nu_{n}\right)}\\
d_{n} & =\left\langle \nu_{n}^{2}\right\rangle _{\psi_{n}\left(\nu_{n}\right)}-\left\langle \nu_{n}\right\rangle _{\psi_{n}\left(\nu_{n}\right)}^{2}
\end{cases}\qquad n\in\left\{ 1,\ldots,N\right\} \label{eq:MeanVarFistLocal}
\end{equation}

We estimate the marginal posteriors from the distribution

\begin{eqnarray}
Q\left(\boldsymbol{\mathbf{\nu}}|\boldsymbol{b}\right) & = & \frac{1}{Z_{Q}}e^{-\frac{\beta}{2}\left(\mathbf{S}\boldsymbol{\nu}-\boldsymbol{b}\right)^{T}\left(\mathbf{S}\boldsymbol{\nu}-\boldsymbol{b}\right)}\prod_{n=1}^{N}\phi_{n}\left(\nu_{n};a_{n},d_{n}\right)\label{eq:GaussPosterior}\\
Z_{Q} & = & \int d^{N}\boldsymbol{\nu}e^{-\frac{\beta}{2}\left(\mathbf{S}\boldsymbol{\nu}-\boldsymbol{b}\right)^{T}\left(\mathbf{S}\boldsymbol{\nu}-\boldsymbol{b}\right)}\prod_{n=1}^{N}\phi_{n}\left(\nu_{n};a_{n},d_{n}\right)\label{eq:ZetaQ}
\end{eqnarray}

Notice that in this approximation fluxes result unbounded. Marginals
obtained by this strategy against the Hit-and-Run Monte Carlo estimate
are shown is Figure \ref{fig:EcoliLocal} (cyan line) for 9 representative
metabolic fluxes of one of the standard model for red blood cell \cite{wiback_monte_2004}.
Marginals evaluated with this simple non-adaptive strategy differ
significantly from the ones evaluated with the Montecarlo sampling
technique. In the following we will show how we can overcome this
limitation by choosing different values for the means $\boldsymbol{a}$
and the variances $\boldsymbol{d}$ in Eq. \ref{eq:GaussPosterior}
making use of the Expectation Propagation algorithm.

\subsection*{Expectation Propagation}

Expectation Propagation (EP) \cite{minka_expectation_2001} is an
efficient technique to approximate intractable (\emph{i.e.} impossible
or impractical to compute analytically) posterior probabilities. EP
was first introduced in the framework of statistical physics as an
advanced mean-field method \cite{opper_gaussian_2000,opper_adaptive_2001}
and further developed for Bayesian inference problems in the seminal
work of Minka \cite{minka_expectation_2001}. 

Let us consider the $n^{th}$ flux and its corresponding approximate
prior $\phi_{n}\left(\nu_{n};a_{n},d_{n}\right)$. We define a \textit{tilted}
distribution $Q^{\left(n\right)}$as

\begin{equation}
Q^{\left(n\right)}\left(\boldsymbol{\nu}|\boldsymbol{b}\right)\equiv\frac{1}{Z_{Q^{\left(n\right)}}}e^{-\frac{\beta}{2}\left(\mathbf{S}\boldsymbol{\nu}-\boldsymbol{b}\right)^{T}\left(\mathbf{S}\boldsymbol{\nu}-\boldsymbol{b}\right)}\psi_{n}\left(\nu_{n}\right)\prod_{m\neq n}\phi_{m}\left(\nu_{m}\right)\label{eq:Q(n)EP}
\end{equation}
 The important difference between the tilted distribution and the
multivariate Gaussian $Q\left(\boldsymbol{\nu}|\boldsymbol{b}\right)$
is that all the intractable priors are approximated as Gaussian probability
densities except for the $n^{th}$ prior which is treated \textit{exactly}.
For this reason we expect that this distribution will be more accurate
than $Q\left(\boldsymbol{\nu}|\boldsymbol{b}\right)$ regarding the
estimate of the statistics of flux $n$ without significantly affecting
the computation of expectations. Bearing in mind that it is a large
number of \emph{exact }priors (\emph{i.e.} the distributions $\{\psi_{i}\}_{i=1,\cdots,N}$)
that make the computation intractable and not a single one, we have
introduced only one exact \emph{intractable} prior in $Q^{\left(n\right)}$.

One way of determining the unknown parameters $a_{n}$ and $d_{n}$
of $\phi_{n}\left(\nu_{n};a_{n},d_{n}\right)$ is to require that
the multivariate Gaussian distribution $Q\left(\boldsymbol{\nu}|\boldsymbol{b}\right)$
is as close as possible to the auxiliary distribution $Q^{\left(n\right)}\left(\boldsymbol{\nu}|\boldsymbol{b}\right)$.
Intuitively, there are at least two possibilities to enforce this
similarity: (i) matching the first and the second moments of the two
distributions (ii) minimizing the Kullback-Leibler divergence $D_{KL}\left(Q^{n}\Vert Q\right)$;
these two methods coincide (see details in Supplementary Note 1).
Thus, we aim at imposing the following moment matching conditions:

\begin{equation}
\begin{cases}
\left\langle \nu_{n}\right\rangle _{Q^{\left(n\right)}}= & \left\langle \nu_{n}\right\rangle _{Q}\\
\left\langle \nu_{n}^{2}\right\rangle _{Q^{\left(n\right)}}= & \left\langle \nu_{n}^{2}\right\rangle _{Q}
\end{cases}\label{eq:mom_match}
\end{equation}
from which we get a relation for the parameters $a_{n}$, $d_{n}$
that is explicitly reported in Section \ref{subsec:Update-rule}. 

EP consists in sequentially repeating this update step for all the
other fluxes and iterate until we reach a numerical convergence. Further
technical details about the convergence are reported in Subsection
\ref{subsec:Numerical-results}. At the fixed point we directly estimate
the marginal posteriors $P_{n}\left(\nu_{n}|\boldsymbol{b}\right)$,
for $n\in\left\{ 1,\ldots,N\right\} $, from marginalization of the
tilted distribution $Q^{\left(n\right)}$ that turns out to be a truncated
Gaussian density in the interval $\left[\nu_{n}^{inf},\nu_{n}^{sup}\right]$
(see Supplementary Note 2). 

At difference from the non-adaptive approach, the EP algorithm determines
the approximated prior density by trying to reproduce the effect that
the true prior density has on variable $\nu_{n}$, including the interaction
of this term with the rest of the system. First, the information encoded
in the stoichiometric matrix is surely encompassed in the computation
of the means and the variances of the approximation since both the
distributions $Q^{\left(n\right)}$ and $Q$ contain the exact expression
of the likelihood. Secondly, the refinement of each prior also depends
on the parameters of \textit{all} the other fluxes.

\begin{figure}
\includegraphics[width=\linewidth]{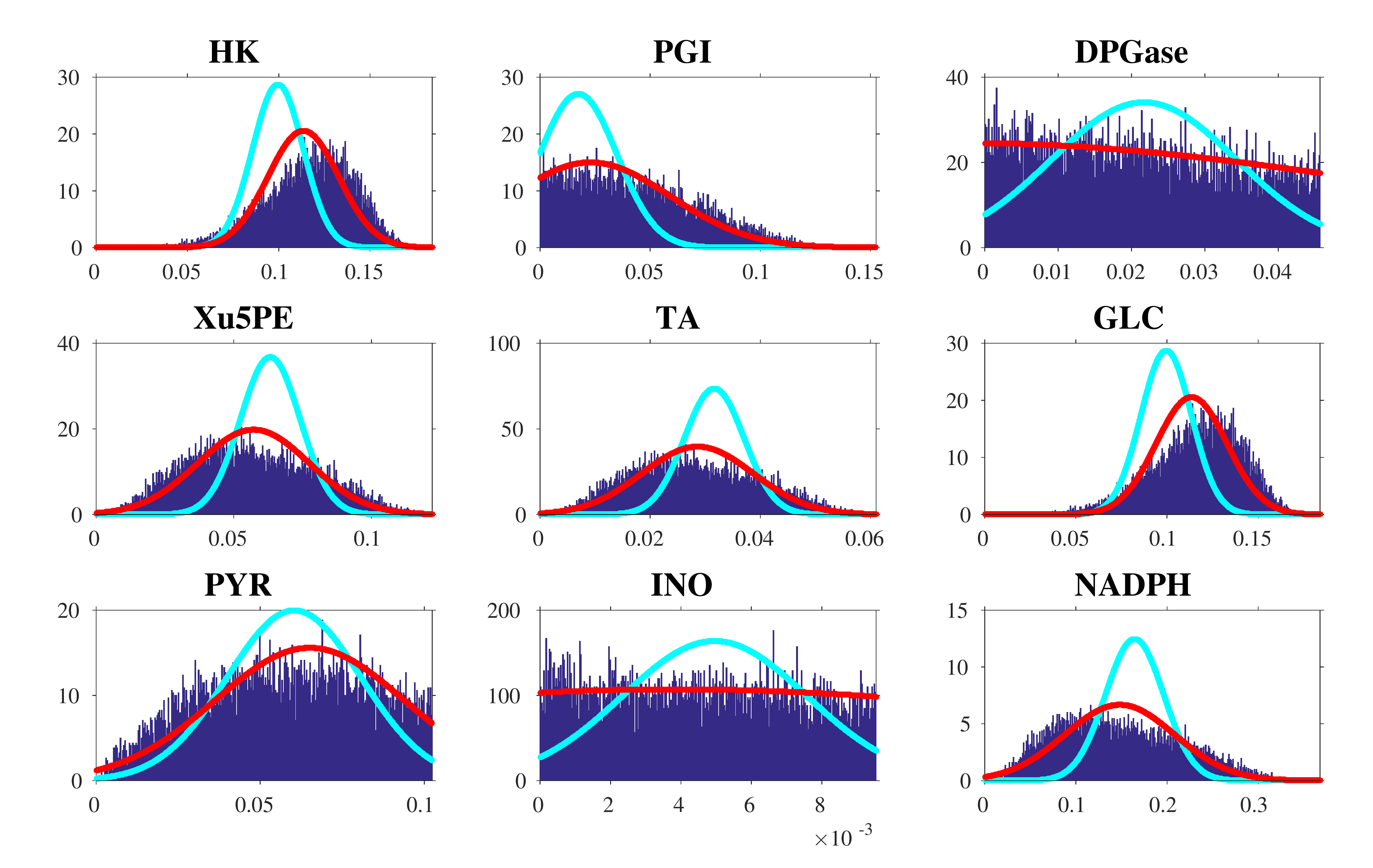}
\caption{Marginal probability densities of nine fluxes of the red blood cell.
The blue bars represent the result of Monte Carlo estimate for $T\sim10^{8}$
sampling points. The cyan line is the result of the non-adaptive Gaussian
approximation while the red line represents the Expectation Propagation
estimate. \label{fig:EcoliLocal}}
\end{figure}

As an example of the accuracy of this technique we report in Figure
\ref{fig:EcoliLocal} (red line) the nine best marginals estimated
by EP of the red blood cell against the results of Hit-and-Run Monte
Carlo sampling. Fig. \ref{fig:EcoliLocal} suggests that this technique
leads to a significant improvement of the non-adaptive approximation
as the plot shows a very good overlap between the distributions provided
by HR and EP. The entire set of marginals and a comparison with a
state-of-the-art message passing algorithm \cite{fernandez-de-cossio-diaz_fast_2016}
is reported in the Supplementary Fig. 2.

\subsection*{Numerical results for large scale metabolic networks \label{subsec:Numerical-results}}

This section is devoted to compare the results of our algorithm against
the outcomes of a state-of-the-art Hit-and-Run Monte Carlo sampling
technique on three representative models of metabolic networks, precisely
the \textit{iJR904} \cite{reed_expanded_2003-1}, the \textit{CHOLnorm
\cite{lewis_large-scale_2010}} and the \textit{RECON1} \cite{duarte_global_2007}
models for \textit{Escherichia Coli}, the \textit{Cholinergic neuron}
and \textit{Homo Sapiens} organisms respectively. In Supplementary
Fig. 3 we report results for a larger set of models all selected from
the Bigg Models database \cite{king2016bigg}. 

Experiments are performed as follows. First we preprocess the stoichiometric
matrix of the model in order to remove all reactions involving metabolites
that are only produced or only degraded\cite{henry2010high}. 

After the pre-processing, we run HR and EP, both implemented on Matlab
or as Matlab libraries, to the reduced model. Let us explain how the
two methods work. Starting from a point lying on the polytope, HR
iteratively chooses a random direction and collects new samples in
that direction such that they also reside in the solution space. In
this work we use an optimized implementation of HR, called \textit{optGpSampler}
\cite{Marchiori_2014}. Regarding the HR simulations we set the number
of sampled points to be equal to $10^{4}$ for an increasing number
of explored configurations $T$ from $10^{4}$ to $10^{7}$ in most
of the cases; for some specific models, i.e. very large networks having
$N\sim10^{3}$ reactions, we explore up to $T\sim10^{9}$ points.
Concerning the EP algorithm we perform the same experiment setting
the $\beta$ parameter to be equal to $10^{10}$ for almost all models.
In only one case (the RECON1 model), we encountered convergence problems
and thus we decreased it to $10^{9}$. Numerical convergence of EP
depends on the refinement of parameters $\boldsymbol{a}$ and $\boldsymbol{d}$
or, more precisely, on the estimate of the marginal distributions
of fluxes. At each iteration $t$ we compute an error $\varepsilon$
which measures how the approximate marginal distributions change in
two consecutive iterations. Formally, we define the error as the maximum
value of the sum of the differences (in absolute values) of the mean
and second moment of the marginal distribution, that is 
\[
\varepsilon^{t}=\max_{n}\left|\left\langle \nu_{n}\right\rangle _{Q^{\left(n\right)}}^{t+1}-\left\langle \nu_{n}\right\rangle _{Q^{\left(n\right)}}^{t}\right|+\left|\left\langle \nu_{n}^{2}\right\rangle _{Q^{\left(n\right)}}^{t+1}-\left\langle \nu_{n}^{2}\right\rangle _{Q^{\left(n\right)}}^{t}\right|.
\]
If $\varepsilon^{t}$ is smaller than a predetermined target precision
(we used $10^{-5})$, the algorithm stops.

To quantitatively compare the two techniques we report here the scatter
plots of variances and means of the approximate marginals computed
via HR and EP. Moreover we estimate the degree of correlation among
the two sets of parameters computing the Pearson product-moment correlation
coefficient 

Notice that we cannot have access to the exact marginals and that
we assume that the results obtained by HR are exact only asymptotically.
Thus our performances, both for the direct comparison of the means
and variances and for the Pearson's coefficient, should be considered
accurate if they are approached by the Monte-Carlo ones for an increasing
number of explored points.

\begin{figure}
\includegraphics[width=\linewidth]{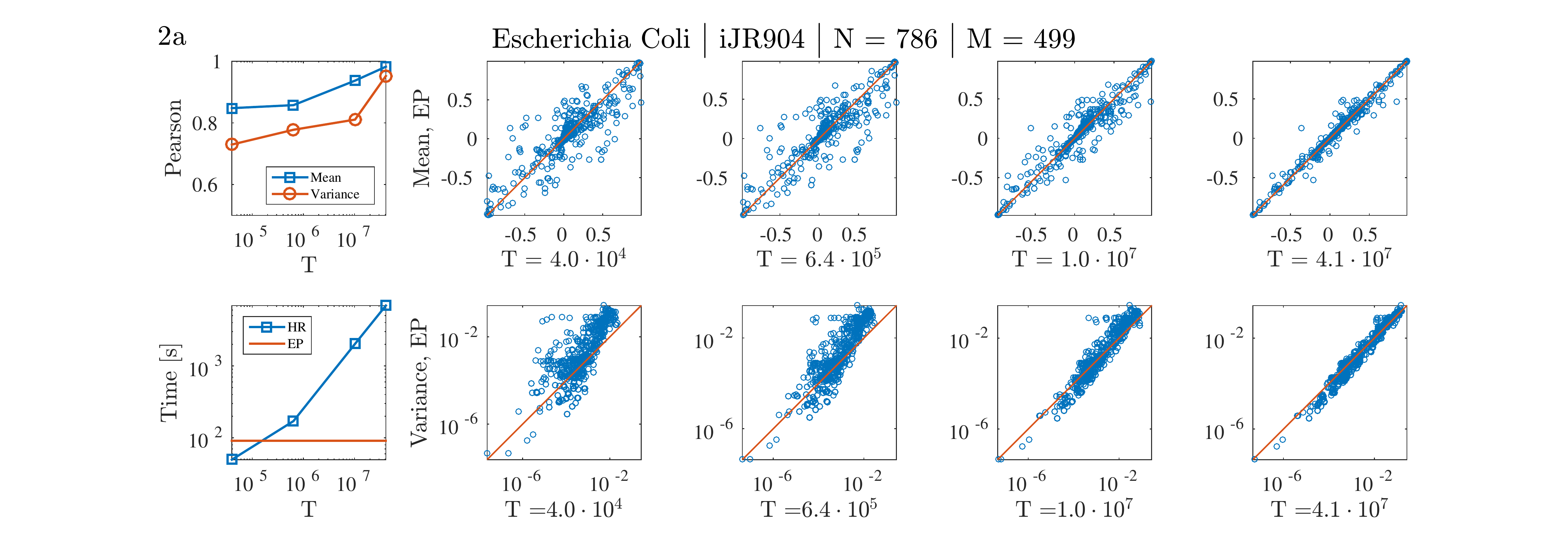}
\includegraphics[width=\linewidth]{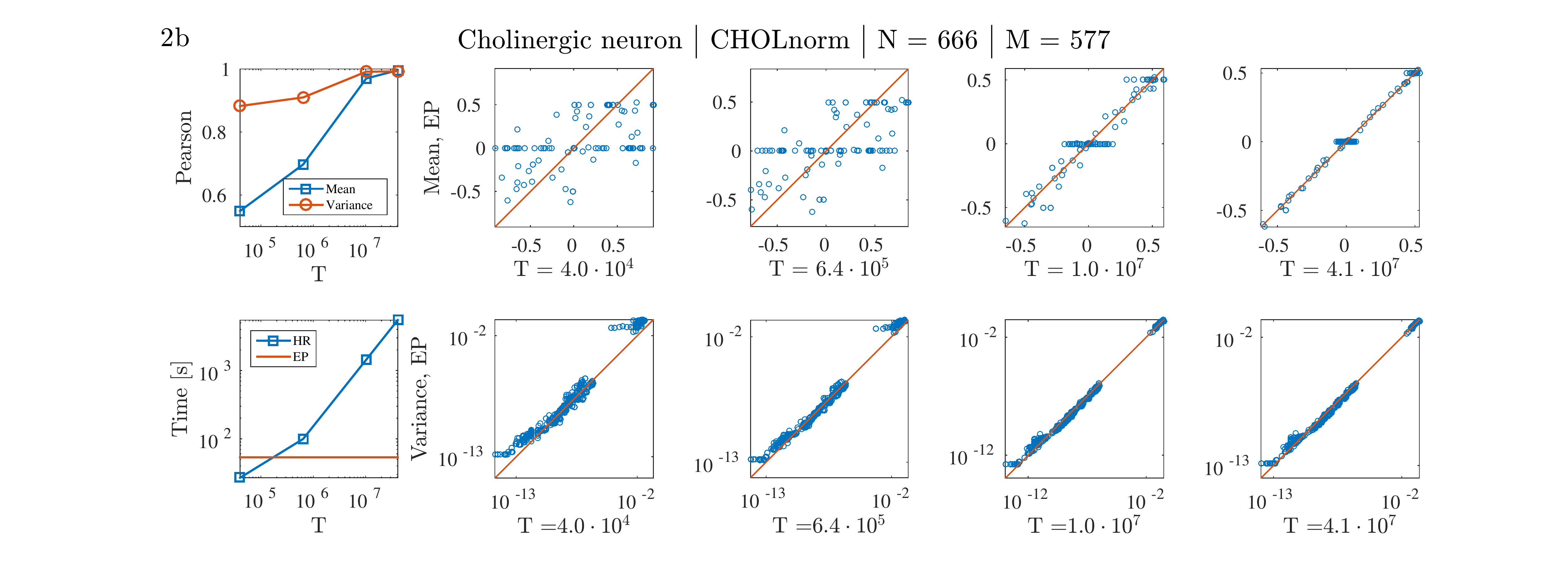}
\includegraphics[width=\linewidth]{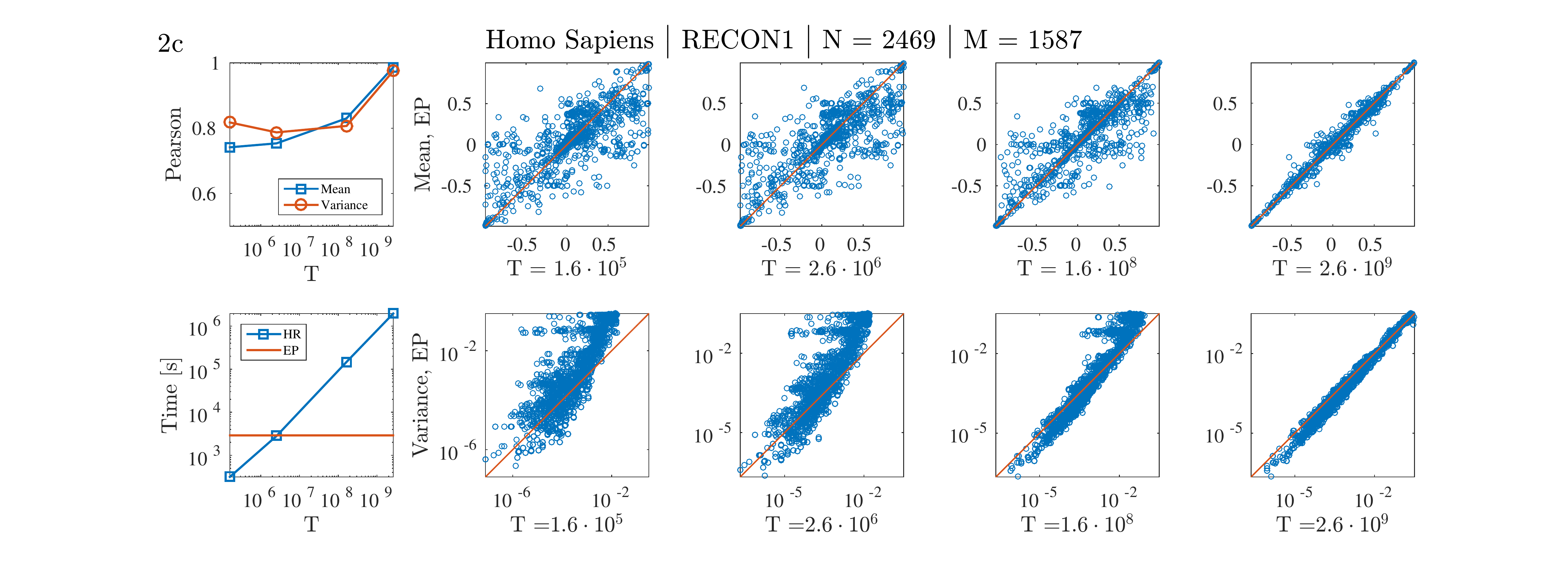}
\caption{Comparison of the results of HR vs EP for \textit{iJR904}, \textit{CHOLnorm}
and \textit{RECON1} models\label{fig:MainRes}. The top-left plot
shows the Pearson correlation coefficients between variances and means
estimated through EP and HR. The bottom-left panel reports the computing
time of EP and HR for different values of \textsl{T}. The plots on
the right are scatter plots of the means and variances of the approximated
marginals computed via EP against the ones estimated via HR for an
increasing number of explored configurations \textsl{T.}}

\end{figure}

The three large subplots in Fig. \ref{fig:MainRes} show the results
for \textit{Escherichia Coli}, \textit{Cholinergic neuron} and \textit{Homo
Sapiens} respectively. For each organism, we report on the top-left
panel the time spent by EP (straight line) and by HR (cyan points)
and on the bottom-left panel the Pearson correlation coefficients.
Both measures of time and correlation, are plotted as functions of
the number of configuration $T$ obtained from the HR algorithm. As
shown in these plots, we can notice that to reach a high correlation
regime a very large number of explored configurations, employing a
computing time that is always several orders of magnitude larger than
the EP running time. This is particularly strinking in the case of
the \textit{RECON1} model, for which we needed to run HR for about
20 days in order to reach results similar to the outcomes of EP, that
converges in less than one hour on the same machine.

To underline how EP seems to approach HR results in the asymptotic
limit, we report in the rest of the sub-figures the scatter plots
of the means (top) and the variances (bottom) of the marginals. On
the $y$-axis we plot the EP means (variances) against the HR means
(variances) for an increasing number of explored configurations, as
indicated in $x$-axis. Results clearly show that as $T$ grows, the
points (both means and variances) are more and more aligned to the
identity line: not only these measures are highly correlated for large
$T$, but they assume very similar values. This is remarkably appreciable
in the results for $CHOLnorm$ model: for $T=4\cdot10^{4}$ the means
of the scatter plots are quite unaligned but as $T$ reaches $4\cdot10^{7}$,
they almost lie on the identity line. In fact, means are poorly correlated
for $T=4\cdot10^{4}$ while the Pearson correlation coefficient is
close to 1 for $T=4\cdot10^{7}$. 

\textbf{}

\subsection*{Study of Escherichia Coli metabolism for a constrained growth rate
\label{subsec:Escherichia-Coli-fluxes}}

The EP formalism can efficiently deal with a slightly modified version
of problem of sampling metabolic networks. Suppose to have access
to experimental measurements of the distribution of some fluxes under
specific environmental conditions. We would like to embed this empirical
knowledge in our algorithm, by matching the posterior distribution
of the measured fluxes with the empirical measurements. Within the
EP scheme, this task corresponds to matching the first two moments
(mean and variance) of the posteriors with the one defined by the
empirical measurements. With the inclusion of empirically established
prior knowledge, we want to investigate how the experimental evidence
is related to the metabolism at the level of reactions or, in other
words, we want to determine how fluxes modify in order to reproduce
the experiments. In this perspective, the EP scheme can easily accommodate
additional constraints on the posteriors by modifying the EP update
equations as outlined in Methods. 

We have tested this variant of EP algorithm on the \textit{iJR904}
model of \textit{Escherichia Coli} for a constrained growth rate.
In fact, one of the few fluxes that are experimentally accessible
is the biomass flux, often measured in terms of doubling per hour.
As a matter of example we decide to extract one of the growth rates
reported in Fig. 3(a) of \cite{PhysRevE.93.012408}; the profile labeled
as \textit{Glc (P5-ori)} can be well fitted by a Gaussian probability
density of mean $0.92\:\mathrm{h^{-1}}$ and variance $0.0324\:\mathrm{h^{-2}}$.
This curve represent single-cell measures of a population of bacteria
living in the so-called \textit{minimal substrate} whose main characteristics
are in principle well caught by the \textit{iJR904} model. We fixed
the bound on the glucose exchange flux EX\_glc(e) such that the maximum
allowed growth rate (about $2\mathrm{\,h^{-1}}$) contained all experimental
values in the profile labeled as \textit{Glc (P5-ori)} of Fig. 3(a)
of \cite{PhysRevE.93.012408}. This was easily computed by fixing
the biomass flux to the desired value and minimizing the glucose exchange
flux using FBA, and gies a the lower bound of the exchanged glucose
flux of $-43\:\mathrm{mmol\,\left(g[DW]\right)^{-1}h^{-1}}$.

\begin{figure}
\includegraphics[width=0.5\linewidth]{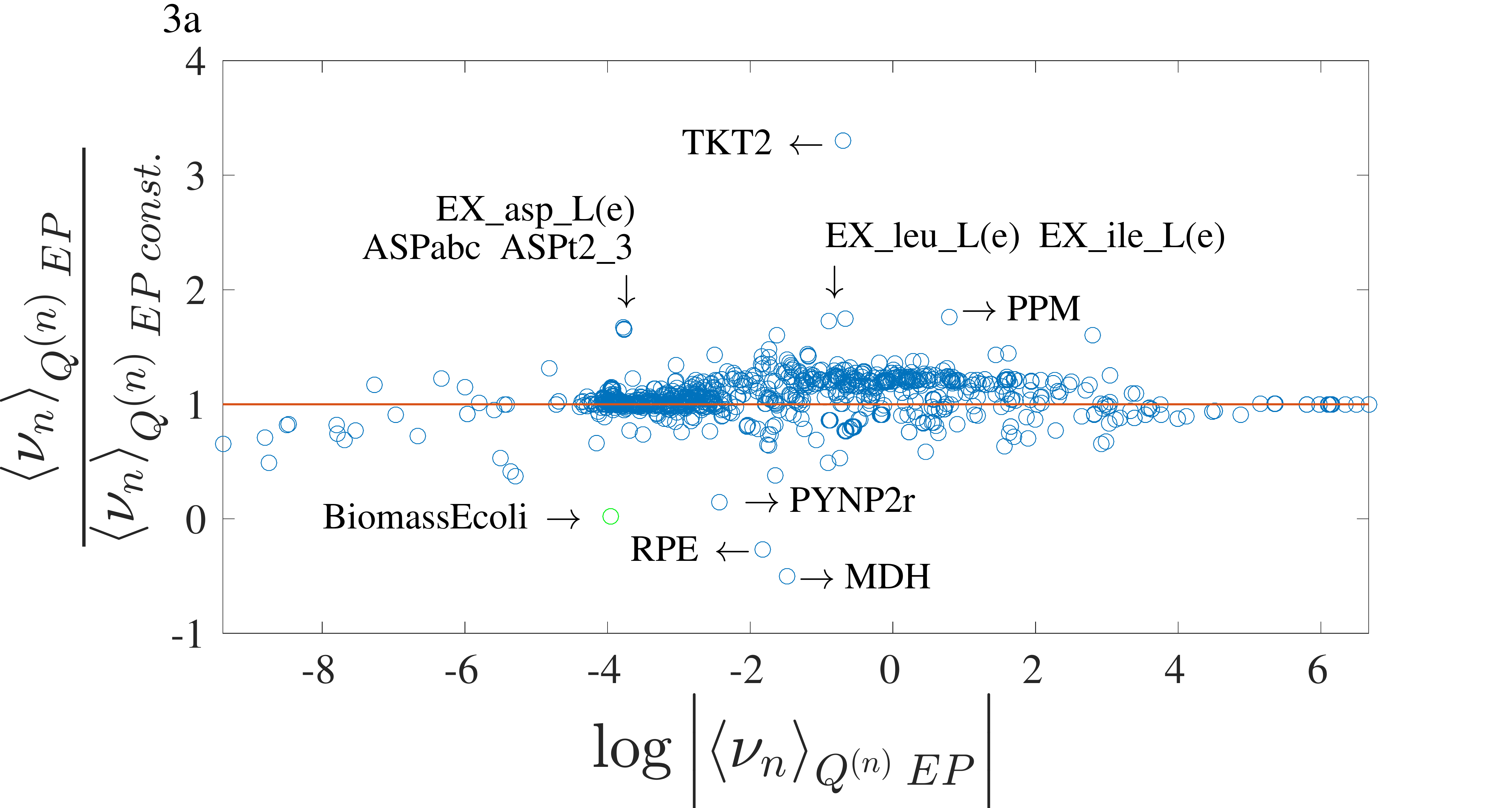}
\includegraphics[width=0.5\linewidth]{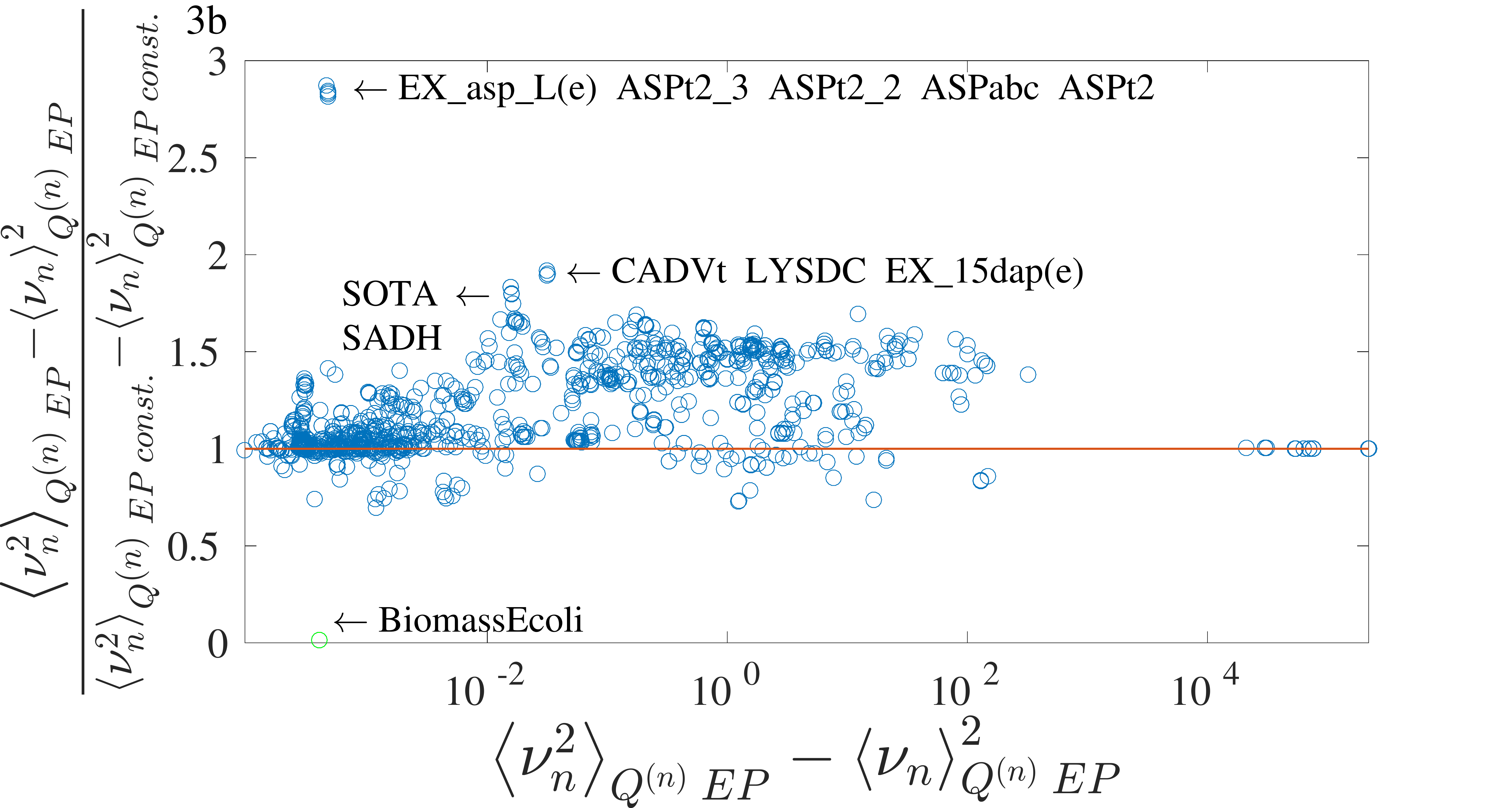}
\caption{Results for a constrained biomass flux. Comparison between the means
(a) and variances (b) of the marginal probability densities for all
the fluxes computed without the additional constraint (unconstrained
case) and with the constrained on the biomass (constrained case).
The green point indicates the biomass flux. \label{fig:Scatter-Const}}
\end{figure}

We then apply EP algorithm to the modified \textit{iJR904} model in
two different conditions. First, we do not impose any additional constraint
and we run the original EP algorithm as described in the previous
section. Then, as described in Methods, we fix the marginal posterior
of the biomass. We can now compare the means and the variances of
all the other fluxes in the two cases and single out those fluxes
that have been more affected by the empirical constraints on the growth
rate. We report in Fig. \ref{fig:Scatter-Const} the plot of the ratio
between the means (Fig. \ref{fig:Scatter-Const} (a)) and the variances
(Fig. \ref{fig:Scatter-Const} (b)) in the unconstrained case and
in the constrained case. In Fig. \ref{fig:Scatter-Const} (a) these
ratios are plotted against the logarithm of the absolute value of
the unconstrained means to differentiate those fluxes having means
close to zero and all the other cases. The ratios of the variances
are instead plotted as a function of the unconstrained variances in
semi-log scale. We can notice that apparently a large fraction of
the fluxes have changed their marginal distribution in order to accommodate
the fixed marginal for the biomass. We have reported the name of the
reactions with the most significant changes; for instance, the marginal
of the \textit{TKT2 }reaction has reduced its mean of more than one
third, while many reactions involving\textit{ aspartate }have significantly
lowered their variances.

To underline the non-trivial results of EP algorithm in the constrained
case, we apply again the standard EP algorithm to the \textit{iJR904}
model when the lower bound and the upper bound of the biomass is fixed
to the average value of the experimental profile. The comparison (not
shown) between the two approaches suggests that the most relevant
change concerns the \textit{EX\_asp\_L(e)} flux as both the average
value and the variance estimated in the second case are about two
times the ones predicted by constrained EP. The distributions of most
other fluxes remain do not considerably change. We underline that
the different behavior of the marginals in the two cases, even if
not significant for most of the fluxes, was in principle unpredictable
without the use of constrained EP; and we do not exclude that fixing
other empirical profiles can lead to very different results. Likewise,
it seems unlikely that the results computed with constrained EP could
be obtained using unbiased samples as provided by standard HR implementations
(see a discussion in Supplementary Note 6).

\section*{Discussion}

In this work we have shown how to study the space of feasible configurations
of metabolic fluxes within a cell via an analytic description of the
marginal probability distribution characterizing each flux. Such marginals
are described as truncated Gaussian densities whose parameters are
determined through an iterative and extremely efficient algorithm,
the Expectation Propagation algorithm. We have compared our predictions
against the estimates provided by HR sampling technique and results
shown in Subsection \ref{subsec:Numerical-results} suggest a very
good agreement between EP and HR for a large number of explored configurations,
$T$. First of all, the direct comparison of the means and variances
of EP vs HR reported in the scatter plots shows that the more we increment
the HR points, the more the scatter points are aligned. Secondly we
see an increment of the correlation between EP and HR statistics for
an increasing number of sampled points; correlations reach values
very close to $1$ for large values of $T$ and for almost all the
models we have considered. The most important point is that the computation
times of EP, at high correlation regime, are always orders of magnitude
lower than HR sampling times. This is extremely time-saving when we
deal with very large networks, as the \textit{RECON1} model for \textit{Homo
Sapiens} where the running time (in seconds) of EP is three order
of magnitude smaller then HR. We underline that exact marginals are
generally inaccessible and we cannot compare our results against a
ground-truth; our measures well approximate ``true'' distributions
as long as the exactness of HR in the asymptotic limit is correct.

We have shown how to include empirical knowledge on distribution of
fluxes on the EP algorithm without compromising the computing time.
More precisely we have investigated how fixing an experimental profile
of the growth rate into the \textit{iJR904} model of Escherichia Coli
affect non-trivially all other fluxes. This is a remarkable advantage
of the EP algorithm with respect to other methods.

EP provides an analytic estimate of each single flux marginal which
relies on the optimization of two parameters, the mean and the variance
of a Gaussian distribution. The formalism allows in principle more
complicated parametrizations of posteriors to include other biological
insights.

EP equations are extremely easy to derive and to implement, as the
main loop can be written in few lines of Matlab code. The method is
iterative, and the number of operations in each iteration scales as
$\Theta\left(N^{3}\right)$, rendering EP extremely convenient in
terms of computation time with respect to existing alternatives.

An shown in Fig. \ref{fig:MainRes} in real cases variances of the
marginal distributions can span several orders of magnitude. This
range of variability implies that also the variances of the approximation
need to allow both very small and huge values. To cope with the numeric
problems that may arise, we allow parameters $\boldsymbol{d}$ to
vary in a finite range of values with the drawback of limiting the
set of allowed Gaussian densities of the approximation. For instance,
a flat distribution cannot be perfectly approximated through a Gaussian
whose variance cannot be arbitrary large; in the opposite extreme,
imposing a lower bound on variances prevents the approximation of
posteriors that are too concentrated on a single point. Thus this
range needs to be reasonably designed in order to catch as many ``true''
variances as possible. In this work we have tried to impose a very
large range of values, typically $\left[10^{-50},\:10^{50}\right]$,
to include as many distributions as possible without compromising
the convergence of the algorithm. Moreover, the Gaussian profile itself
is surely a limitation of the approximation as true marginals can
have in principle arbitrary profiles. 

EP performances are sensitive to the parameter $\beta$ and equations
become numerically unstable for too large $\beta$ (e.g. $10^{11}-10^{12}$).
On the other hand $\beta$ can be seen as the inverse-variance of
a Gaussian noise affecting the conservation laws. The nature of this
noise could depend on localization properties on the cell and real
thermal noise. In this case, an optimization of the free energy with
respect to $\beta$ can in principle lead to better predictions.

\section*{Methods}

\subsection*{Update rule \label{subsec:Update-rule}}

The algorithm described in the Expectation Propagation section relies
on local moves in which, at each step, we refine only the parameters
of \textit{one} single prior, minimizing the dissimilarity between
the auxiliary $tilted$ distribution $Q^{\left(n\right)}$ and $Q$.
The values of the mean and of the variance of $\phi_{n}\left(\nu_{n};a_{n},d_{n}\right)$
are iteratively tuned in a way that the first and second moments of
the two distributions match. The update rule for the parameters $a_{n}$
and $d_{n}$ of the Gaussian prior will be derived in details in the
following section.

Let us express the auxiliary density $Q^{\left(n\right)}$in Eq. \ref{eq:Q(n)EP}
as a standard multivariate Gaussian distribution times the exact prior
of the $n^{th}$ flux as

\begin{eqnarray}
Q^{\left(n\right)}\left(\boldsymbol{\nu}|\boldsymbol{b}\right) & = & \frac{1}{Z_{Q^{\left(n\right)}}}e^{-\frac{\beta}{2}(\mathbf{S}\boldsymbol{\nu}-\boldsymbol{b})^{T}(\mathbf{S}\boldsymbol{\nu}-\boldsymbol{b})-\frac{1}{2}(\boldsymbol{\nu}-\boldsymbol{a})^{T}\mathbf{D}(\boldsymbol{\nu}-\boldsymbol{a})}\psi_{n}\left(\nu_{n}\right)\\
 & = & \frac{1}{\tilde{Z}_{Q^{\left(n\right)}}}e^{-\frac{1}{2}(\boldsymbol{\nu}-\boldsymbol{\bar{\nu}})^{T}\boldsymbol{\Sigma}^{-1}(\boldsymbol{\nu}-\boldsymbol{\bar{\nu}})}\psi_{n}\left(\nu_{n}\right)\label{eq:QnSigma}
\end{eqnarray}
where $\tilde{Z}_{Q^{\left(n\right)}}=Z_{Q^{\left(n\right)}}e^{\frac{\beta}{2}\boldsymbol{b}^{T}\boldsymbol{b}-\frac{1}{2}\boldsymbol{\bar{\nu}}^{T}\boldsymbol{\bar{\nu}}}$,
$\mathbf{D}$ is a diagonal matrix with components $D_{mm}=\frac{1}{d_{m}}$
for $m\neq n$ and $D_{nn}=0$ (and of course non-diagonal terms $D_{mk}=0$
if $m\neq k$). The covariance matrix $\boldsymbol{\Sigma}^{-1}$
and the mean vector $\bar{\boldsymbol{\nu}}$ satisfy:

\begin{equation}
\begin{cases}
\boldsymbol{\Sigma}^{-1} & =\beta\mathbf{S}^{T}\mathbf{S}+\mathbf{D}\\
\bar{\boldsymbol{\nu}} & =\boldsymbol{\Sigma}\left(\beta\mathbf{S}^{t}\boldsymbol{b}+\mathbf{D}\boldsymbol{a}\right)
\end{cases}\label{eq:sigma_nubar}
\end{equation}

Note that we are omitting for notational simplicity the dependence
of $\mathbf{D},\boldsymbol{\Sigma},\overline{\boldsymbol{\nu}}$ on
$n$. Equivalently
\begin{equation}
Q\left(\boldsymbol{\nu}|\boldsymbol{b}\right)=\frac{1}{\tilde{Z}_{Q}}e^{-\frac{1}{2}(\boldsymbol{\nu}-\boldsymbol{\bar{\nu}})^{T}\boldsymbol{\Sigma}^{-1}(\boldsymbol{\nu}-\boldsymbol{\bar{\nu}})}\phi_{n}(\nu_{n};a_{n},d_{n})
\end{equation}
where $\tilde{Z}_{Q}=Z_{Q}e^{\frac{\beta}{2}\boldsymbol{b}^{T}\boldsymbol{b}-\frac{1}{2}\boldsymbol{\bar{\nu}}^{T}\boldsymbol{\bar{\nu}}}$.
If we now exploit the moment matching condition in Eq. \eqref{eq:mom_match}
(a detailed calculation of the moments of $Q$ and $Q^{\left(n\right)}$
expressed in standard form is reported in Supplementary Notes 2-3)
we obtain an update equation for the mean and the variance:

\begin{equation}
\begin{cases}
d_{n} & =\left(\frac{1}{\left\langle \nu_{n}^{2}\right\rangle _{Q^{\left(n\right)}}-\left\langle \nu_{n}\right\rangle _{Q^{\left(n\right)}}^{2}}-\frac{1}{\Sigma_{nn}}\right)^{-1}\\
a_{n} & =d_{n}\left[\left\langle \nu_{n}\right\rangle _{Q^{\left(n\right)}}\left(\frac{1}{d_{n}}+\frac{1}{\Sigma_{nn}}\right)-\frac{\bar{\nu}_{n}}{\Sigma_{nn}}\right]
\end{cases}\label{eq:update_a_d}
\end{equation}

Notice that the sequential update scheme described in the Expectation
Propagation section requires the inversion of the matrix $\boldsymbol{\Sigma}^{-1}$
each time that we have to refine the parameters of flux $n$, leading
to $N$ inversions per iteration amounting to $\Theta\left(N^{4}\right)$
operations per iteration. We propose in Supplementary Note 4 a parallel
update, that needs only one matrix inversion per iteration, i.e. $\Theta\left(N^{3}\right)$
operations per iteration.

\subsection*{Update equations for a constrained posterior \label{sec:Exp-Estimation}}

Let us assume to have access to experimental measures of the (marginal)
posterior $f\left(\nu_{i}\right)$ for flux $i$. We aim at determining
how the posteriors of other fluxes would modify to fit with the experimental
results compared, for instance, to the unconstrained case. The so-called
\emph{maximum entropy principle} \cite{jaynes1957information} dictates
that the most unconstrained distribution which is consistent with
the experiment, prior distributions and flux conservation $\mathbf{S}\boldsymbol{\nu}=\boldsymbol{b}$,
is simply 
\begin{equation}
P\left(\boldsymbol{\nu}|\boldsymbol{b}\right)=\frac{1}{Z}e^{-\frac{\beta}{2}\left(\mathbf{S}\boldsymbol{\nu}-\boldsymbol{b}\right)^{T}\left(\mathbf{S}\boldsymbol{\nu}-\boldsymbol{b}\right)}\prod_{n=1}^{N}\mathbb{I}\left(\nu_{n}\in\left[\nu_{n}^{inf},\nu_{n}^{sup}\right]\right)g\left(\nu_{i}\right)\,\,\,\,,\label{eq:lagrange}
\end{equation}
where $\beta\to\infty$ and $g\left(\nu_{i}\right)$ is the (exponential
of the) function of unknown Lagrange multipliers that has to be determined
in order for the constraint $\int\prod_{n\neq i}d\nu_{n}P\left(\boldsymbol{\nu}|\boldsymbol{b}\right)=f\left(\nu_{i}\right)$
to be satisfied. In the particular case in which the posterior can
be reasonably fitted by a Gaussian distribution $\mathcal{N}\left(\nu_{i}|a_{i}^{exp},d_{i}^{exp}\right)$,
then it suffices to consider also a Gaussian $g\left(\nu_{i}\right)=\mathcal{N}\left(\nu_{i}|a_{i},d_{i}\right)=\phi_{i}\left(\nu_{i}\mid a_{i},d_{i}\right)$
with only two free parameters. The determination of $a_{i},d_{i}$
can be achieved by slightly modifying the EP update for flux $i$.
Assuming as before that the prior of each flux $n\neq i$ can be approximated
as a Gaussian profile $\phi_{n}\left(\nu_{n};a_{n},d_{n}\right)$
of parameters $a_{n}$ and $d_{n}$, also to be determined, we must
impose that 

\begin{align}
\mathcal{N}\left(\nu_{i}|a_{i}^{exp},d_{i}^{exp}\right) & \propto\mathcal{N}\left(\nu_{i}|a_{i},d_{i}\right)\int\prod_{n\neq i}d\nu_{n}Q\left(\boldsymbol{\nu}|\boldsymbol{b}\right)\\
 & \propto\phi_{i}\left(\nu_{i};a_{i},d_{i}\right)e^{-\frac{\left(\nu_{i}-\bar{\nu}_{i}\right)^{2}}{2\Sigma_{ii}}}\label{eq:prior_estimate}
\end{align}
where the distribution $Q\left(\boldsymbol{\nu}|\boldsymbol{b}\right)$
is the one in Eq. \eqref{eq:GaussPosterior}. Since both the left-hand
side and the right-hand side of Eq. \eqref{eq:prior_estimate} contain
Gaussian distributions, the relations for $a_{i}$ and $d_{i}$ can
be easily computed and take the form 

\begin{equation}
\begin{cases}
d_{i} & =\left(\frac{1}{d_{i}^{exp}}-\frac{1}{\Sigma_{ii}}\right)^{-1}\\
a_{i} & =d_{i}\left(\frac{a_{i}^{exp}}{d_{i}^{exp}}-\frac{\bar{\nu}_{i}}{\Sigma_{ii}}\right)
\end{cases}
\end{equation}

This expression is exactly the same in Eq. \eqref{eq:update_a_d}
if we replace the mean and the variance of the tilted distribution
with the experimental ones. 

\subsection*{Technical details}

The computations were performed on a Dell Poweredge server with 128
Gb of memory and 48 AMD Opteron cpus clocked at 1.9Ghz. No constraint
have been placed on the number of cpu threads, allowing both EP and
HR to parallelize their processes. We observed that EP used 2-3 cores,
exclusively in the matrix inversion phase (which was time-dominant),
while HR employed a variable number of cores (around 6 or 7 at some
times). For this reason only the order of magnitude of computing times
of HR and EP are fairly comparable but they are sufficient to underline
the differences between the two algorithms.

\subsection*{Data and code availability}

All data generated or analysed during this study are included in this
published article (and its Supplementary Information file). 

An implementation of the algorithm presented in this work is publicly
available at \url{https://github.com/anna-pa-m/Metabolic-EP}.

Authors declare no competing interest.

\section*{Acknowledgment}

We warmly thank Andrea De Martino, Roberto Mulet, Jorge Fernández
de Cossio, Eduardo Martinez Montes for interesting discussions. We
acknowledge support from project SIBYL, financed by Fondazione CRT
under the initiative ``La ricerca dei Talenti'' and project ``From
cellulose to biofuel through Clostridium cellulovorans: an alternative
biorefinery approach'' of University of Turin, financed by Compagnia
di Sanpaolo.

\section*{Authors contribution}

All authors contributed equally to this work.

\bibliographystyle{unsrt}

\begin{thebibliography}{10}

\bibitem{nelson2008lehninger}
David~L Nelson, Albert~L Lehninger, and Michael~M Cox.
\newblock {\em Lehninger principles of biochemistry}.
\newblock Macmillan, 2008.

\bibitem{palsson2015systems}
Bernhard~{\O} Palsson.
\newblock {\em Systems biology: constraint-based reconstruction and analysis}.
\newblock Cambridge University Press, 2015.

\bibitem{varma_metabolic_1994}
Amit Varma and Bernhard~O. Palsson.
\newblock Metabolic {Flux} {Balancing}: {Basic} {Concepts}, {Scientific} and
  {Practical} {Use}.
\newblock {\em Nat Biotech}, 12(10):994--998, October 1994.

\bibitem{kauffman_advances_2003}
Kenneth~J Kauffman, Purusharth Prakash, and Jeremy~S Edwards.
\newblock Advances in flux balance analysis.
\newblock {\em Current Opinion in Biotechnology}, 14(5):491--496, October 2003.

\bibitem{Suthers2007}
Patrick~F. Suthers, Anthony~P. Burgard, Madhukar~S. Dasika, Farnaz Nowroozi,
  Stephen~Van Dien, Jay~D. Keasling, and Costas~D. Maranas.
\newblock Metabolic flux elucidation for large-scale models using 13c labeled
  isotopes.
\newblock {\em Metabolic Engineering}, 9(5-6):387 -- 405, 2007.

\bibitem{Marchiori_2014}
Wout Megchelenbrink, Martijn Huynen, and Elena Marchiori.
\newblock {\it optGpSampler}: An improved tool for uniformly sampling the
  solution-space of genome-scale metabolic networks.
\newblock {\em PLoS ONE}, 9(2):e86587, 02 2014.

\bibitem{fernandez-de-cossio-diaz_fast_2016}
J.~Fernandez-de Cossio-Diaz and R.~Mulet.
\newblock Fast inference of ill-posed problems within a convex space.
\newblock {\em Journal of Statistical Mechanics: Theory and Experiment},
  2016(7):073207, 2016.

\bibitem{DeMartinoParisi2014}
Daniele De~Martino, Matteo Mori, and Valerio Parisi.
\newblock Uniform sampling of steady states in metabolic networks:
  Heterogeneous scales and rounding.
\newblock {\em PLoS ONE}, 10(4):e0122670, 04 2015.

\bibitem{Dyer_Frieze1988}
M.~E. Dyer and A.~M. Frieze.
\newblock On the complexity of computing the volume of a polyhedron.
\newblock {\em SIAM Journal on Computing}, 17(5):967--974, 1988.

\bibitem{Smith1984}
Robert~L. Smith.
\newblock Efficient monte carlo procedures for generating points uniformly
  distributed over bounded regions.
\newblock {\em Operations Research}, 32(6):1296--1308, 1984.

\bibitem{turchin_computation_1971}
V.~Turchin.
\newblock On the {Computation} of {Multidimensional} {Integrals} by the
  {Monte}-{Carlo} {Method}.
\newblock {\em Theory Probab. Appl.}, 16(4):720--724, January 1971.

\bibitem{kaufman1998}
Robert L.~Smith David E.~Kaufman.
\newblock Direction choice for accelerated convergence in hit-and-run sampling.
\newblock {\em Operations Research}, 46(1):84--95, 1998.

\bibitem{Schellenberger2011}
Jan Schellenberger, Richard Que, Ronan M.~T. Fleming, Ines Thiele, Jeffrey~D.
  Orth, Adam~M. Feist, Daniel~C. Zielinski, Aarash Bordbar, Nathan~E. Lewis,
  Sorena Rahmanian, Joseph Kang, Daniel~R. Hyduke, and Bernhard~O. Palsson.
\newblock Quantitative prediction of cellular metabolism with constraint-based
  models: the cobra toolbox v2.0.
\newblock {\em Nat. Protocols}, 6(9):1290--1307, Sep 2011.

\bibitem{Price2004}
Nathan~D. Price, Jan Schellenberger, and Bernhard~O. Palsson.
\newblock Uniform sampling of steady-state flux spaces: Means to design
  experiments and to interpret enzymopathies.
\newblock {\em Biophysical Journal}, 87(4):2172 -- 2186, 2004.

\bibitem{schellenberger_use_2009}
Jan Schellenberger and Bernhard~Ø Palsson.
\newblock Use of {Randomized} {Sampling} for {Analysis} of {Metabolic}
  {Networks}.
\newblock {\em Journal of Biological Chemistry}, 284(9):5457--5461, February
  2009.

\bibitem{TaheriAraghi2015}
Sattar Taheri-Araghi, Serena Bradde, John~T. Sauls, Norbert~S. Hill, Petra~Anne
  Levin, Johan Paulsson, Massimo Vergassola, and Suckjoon Jun.
\newblock Cell-size control and homeostasis in bacteria.
\newblock {\em Current Biology}, 25(3):385 -- 391, 2015.

\bibitem{de_martino_scalable_2012}
Daniele De~Martino, Matteo Figliuzzi, Andrea De~Martino, and Enzo Marinari.
\newblock A {Scalable} {Algorithm} to {Explore} the {Gibbs} {Energy}
  {Landscape} of {Genome}-{Scale} {Metabolic} {Networks}.
\newblock {\em PLoS Comput Biol}, 8(6):e1002562, June 2012.

\bibitem{de_martino_growth_2016}
Daniele De~Martino, Fabrizio Capuani, and Andrea De~Martino.
\newblock Growth against entropy in bacterial metabolism: the phenotypic
  trade-off behind empirical growth rate distributions in {E}. coli.
\newblock {\em arXiv preprint arXiv:1601.03243}, 2016.

\bibitem{Mezard_Montanari_book}
Marc Mezard and Andrea Montanari.
\newblock {\em Information, Physics, and Computation}.
\newblock Oxford University Press, Inc., New York, NY, USA, 2009.

\bibitem{braunstein_estimating_2008}
Alfredo Braunstein, Roberto Mulet, and Andrea Pagnani.
\newblock Estimating the size of the solution space of metabolic networks.
\newblock {\em BMC Bioinformatics}, 9(1):240, May 2008.

\bibitem{braunstein_space_2008}
A.~Braunstein, R.~Mulet, and A.~Pagnani.
\newblock The space of feasible solutions in metabolic networks.
\newblock {\em J. Phys.: Conf. Ser.}, 95(1):012017, January 2008.

\bibitem{massucci2013}
Francesco~Alessandro Massucci, Francesc Font-Clos, Andrea De~Martino, and
  Isaac~P\'erez Castillo.
\newblock A novel methodology to estimate metabolic flux distributions in
  constraint-based models.
\newblock {\em Metabolites}, 3(3):838, 2013.

\bibitem{font-clos_weighted_2012}
Francesc Font-Clos, Francesco~Alessandro Massucci, and Isaac~Pérez Castillo.
\newblock A weighted belief-propagation algorithm for estimating volume-related
  properties of random polytopes.
\newblock {\em Journal of Statistical Mechanics: Theory and Experiment},
  2012(11):P11003, 2012.

\bibitem{orth_reconstruction_2010}
Jeffrey~D. Orth, Bernhard~Ø. Palsson, and R.~M.~T. Fleming.
\newblock Reconstruction and {Use} of {Microbial} {Metabolic} {Networks}: the
  {Core} {Escherichia} coli {Metabolic} {Model} as an {Educational} {Guide}.
\newblock {\em EcoSal Plus}, 4(1), September 2010.

\bibitem{wiback_monte_2004}
Sharon~J. Wiback, Iman Famili, Harvey~J. Greenberg, and Bernhard~Ø Palsson.
\newblock Monte {Carlo} sampling can be used to determine the size and shape of
  the steady-state flux space.
\newblock {\em Journal of Theoretical Biology}, 228(4):437--447, June 2004.

\bibitem{minka_expectation_2001}
Thomas~P. Minka.
\newblock Expectation propagation for approximate {Bayesian} inference.
\newblock In {\em Proceedings of the {Seventeenth} conference on {Uncertainty}
  in artificial intelligence}, pages 362--369. Morgan Kaufmann Publishers Inc.,
  2001.

\bibitem{opper_gaussian_2000}
M.~Opper and O.~Winther.
\newblock Gaussian processes for classification: mean-field algorithms.
\newblock {\em Neural Computation}, 12(11):2655--2684, November 2000.

\bibitem{opper_adaptive_2001}
Manfred Opper and Ole Winther.
\newblock Adaptive and self-averaging {Thouless}-{Anderson}-{Palmer} mean-field
  theory for probabilistic modeling.
\newblock {\em Physical Review E}, 64(5):056131, October 2001.

\bibitem{reed_expanded_2003-1}
Jennifer~L. Reed, Thuy~D. Vo, Christophe~H. Schilling, and Bernhard~O. Palsson.
\newblock An expanded genome-scale model of {Escherichia} coli {K}-12 ({iJR}904
  {GSM}/{GPR}).
\newblock {\em Genome Biology}, 4(9):R54, 2003.

\bibitem{lewis_large-scale_2010}
Nathan~E. Lewis, Gunnar Schramm, Aarash Bordbar, Jan Schellenberger, Michael~P.
  Andersen, Jeffrey~K. Cheng, Nilam Patel, Alex Yee, Randall~A. Lewis, Roland
  Eils, Rainer König, and Bernhard~Ø Palsson.
\newblock Large-scale in silico modeling of metabolic interactions between cell
  types in the human brain.
\newblock {\em Nature Biotechnology}, 28(12):1279--1285, December 2010.

\bibitem{duarte_global_2007}
Natalie~C. Duarte, Scott~A. Becker, Neema Jamshidi, Ines Thiele, Monica~L. Mo,
  Thuy~D. Vo, Rohith Srivas, and Bernhard~Ø. Palsson.
\newblock Global reconstruction of the human metabolic network based on genomic
  and bibliomic data.
\newblock {\em Proceedings of the National Academy of Sciences of the United
  States of America}, 104(6):1777--1782, February 2007.

\bibitem{king2016bigg}
Zachary~A King, Justin Lu, Andreas Dr{\"a}ger, Philip Miller, Stephen
  Federowicz, Joshua~A Lerman, Ali Ebrahim, Bernhard~O Palsson, and Nathan~E
  Lewis.
\newblock Bigg models: A platform for integrating, standardizing and sharing
  genome-scale models.
\newblock {\em Nucleic acids research}, 44(D1):D515--D522, 2016.

\bibitem{henry2010high}
Christopher~S Henry, Matthew DeJongh, Aaron~A Best, Paul~M Frybarger, Ben
  Linsay, and Rick~L Stevens.
\newblock High-throughput generation, optimization and analysis of genome-scale
  metabolic models.
\newblock {\em Nature biotechnology}, 28(9):977--982, 2010.

\bibitem{PhysRevE.93.012408}
Andrew~S. Kennard, Matteo Osella, Avelino Javer, Jacopo Grilli, Philippe Nghe,
  Sander~J. Tans, Pietro Cicuta, and Marco Cosentino~Lagomarsino.
\newblock Individuality and universality in the growth-division laws of single
  \textit{E. coli} cells.
\newblock {\em Phys. Rev. E}, 93:012408, Jan 2016.

\bibitem{jaynes1957information}
Edwin~T Jaynes.
\newblock Information theory and statistical mechanics.
\newblock {\em Physical review}, 106(4):620, 1957.

\end{thebibliography}

\pagebreak{}

\section*{Supplementary notes}

\subsection*{Supplementary note 1. KL-divergence minimization of the full conditional
	probabilities}

\renewcommand{\theequation}{S\arabic{equation}}

Let us now rewrite the full probability distributions in Eq. (15)
and Eq. (10) making explicit the dependency of the normalization factors
with respect to the parameters $a_{n}$, $d_{n}$:

\begin{eqnarray}
Q^{\left(n\right)}\left(\boldsymbol{\nu}|\boldsymbol{b}\right) & = & \frac{1}{\tilde{Z}_{Q^{\left(n\right)}}}e^{-\frac{1}{2}(\boldsymbol{\nu}-\boldsymbol{\bar{\nu}})^{T}\boldsymbol{\Sigma}^{-1}(\boldsymbol{\nu}-\boldsymbol{\bar{\nu}})}\psi_{n}(\nu_{n})\\
Q\left(\boldsymbol{\nu}|\boldsymbol{b}\right) & = & \frac{1}{\tilde{Z}_{Q}\left(a_{n},d_{n}\right)}e^{-\frac{1}{2}(\boldsymbol{\nu}-\boldsymbol{\bar{\nu}})^{T}\boldsymbol{\Sigma}^{-1}(\boldsymbol{\nu}-\boldsymbol{\bar{\nu}})}e^{-\frac{(\nu_{n}-a_{n})^{2}}{2d_{n}}}
\end{eqnarray}
where the partition functions are given by:

\begin{eqnarray}
\tilde{Z}_{Q^{\left(n\right)}} & = & \int d^{N}\boldsymbol{\nu}e^{-\frac{1}{2}(\boldsymbol{\nu}-\boldsymbol{\bar{\nu}})^{T}\boldsymbol{\Sigma}^{-1}(\boldsymbol{\nu}-\boldsymbol{\bar{\nu}})}\psi_{n}(\nu_{n})\\
\tilde{Z}_{Q}\left(a_{n},d_{n}\right) & = & \int d^{N}\boldsymbol{\nu}e^{-\frac{1}{2}(\boldsymbol{\nu}-\boldsymbol{\bar{\nu}})^{T}\boldsymbol{\Sigma}^{-1}(\boldsymbol{\nu}-\boldsymbol{\bar{\nu}})}e^{-\frac{(\nu_{n}-a_{n})^{2}}{2d_{n}}}
\end{eqnarray}

Let us compute $D_{KL}(Q^{\left(n\right)}||Q)$:

\begin{eqnarray}
D_{KL}(Q^{\left(n\right)}||Q) & = & \int Q^{\left(n\right)}(\boldsymbol{\nu}|\boldsymbol{b})\log\left(\frac{\psi_{n}\left(\nu_{n}\right)\tilde{Z}_{Q}\left(a_{n},d_{n}\right)}{\phi_{n}\left(\nu_{n}\right)\tilde{Z}_{Q^{\left(n\right)}}}\right)d^{N}\boldsymbol{\nu}\\
& = & \int Q^{\left(n\right)}(\boldsymbol{\nu}|\boldsymbol{b})\log\left(\frac{\tilde{Z}_{Q}\left(a_{n},d_{n}\right)}{e^{-\frac{(\nu_{n}-a_{n})^{2}}{2d_{n}}}}\right)d^{N}\boldsymbol{\nu}+const\\
& = & \int Q^{\left(n\right)}(\boldsymbol{\nu}|\boldsymbol{b})\left[\frac{(\nu_{n}-a_{n})^{2}}{2d_{n}}+\log\tilde{Z}_{Q}\left(a_{n},d_{n}\right)\right]d^{N}\boldsymbol{\nu}+const\\
& = & \frac{\langle(\nu_{n}-a_{n})^{2}\rangle_{Q^{\left(n\right)}}}{2d_{n}}+\log\tilde{Z}_{Q}\left(a_{n},d_{n}\right)+const
\end{eqnarray}
where $const$ does not depend on $a_{n}$ and $d_{n}$. We aim at
minimizing $D_{KL}(Q^{\left(n\right)}||Q)$ with respect to $a_{n},d_{n}$:

\begin{eqnarray}
\frac{\partial D_{KL}(Q^{\left(n\right)}||Q)}{\partial a_{n}} & = & \frac{-\langle\nu_{n}\rangle_{Q^{\left(n\right)}}+a_{n}}{d_{n}}+\frac{1}{\tilde{Z}_{Q}}\frac{\partial\tilde{Z}_{Q}}{\partial a_{n}}\\
\frac{\partial D_{KL}(Q^{\left(n\right)}||Q)}{\partial d_{n}} & = & -\frac{\langle(\nu_{n}-a_{n})^{2}\rangle_{Q^{\left(n\right)}}}{2d_{n}^{2}}+\frac{1}{\tilde{Z}_{Q}}\frac{\partial\tilde{Z}_{Q}}{\partial d_{n}}\label{eq:DKLderiv}
\end{eqnarray}
Since we can move the derivative inside the integration in $\frac{\partial\tilde{Z}_{Q}}{\partial a_{n}}$
and in $\frac{\partial\tilde{Z}_{Q}}{\partial d_{n}}$ we get:

\begin{eqnarray*}
	\frac{1}{\tilde{Z}_{Q}}\frac{\partial\tilde{Z}_{Q}}{\partial a_{n}} & = & \frac{1}{\tilde{Z}_{Q}}\int d^{N}\boldsymbol{\nu}e^{-\frac{1}{2}(\boldsymbol{\nu}-\boldsymbol{\bar{\nu}})^{T}\boldsymbol{\Sigma}^{-1}(\boldsymbol{\nu}-\boldsymbol{\bar{\nu}})}e^{-\frac{\left(\nu_{n}-a_{n}\right)^{2}}{2d_{n}}}\frac{\left(\nu_{n}-a_{n}\right)}{d_{n}}\\
	& = & \langle\frac{\nu_{n}-a_{n}}{d_{n}}\rangle_{Q}
\end{eqnarray*}

\begin{eqnarray*}
	\frac{1}{\tilde{Z}_{Q}}\frac{\partial\tilde{Z}_{Q}}{\partial d_{n}} & = & \frac{1}{Z_{Q}}\int d^{N}\boldsymbol{\nu}e^{-\frac{1}{2}(\boldsymbol{\nu}-\boldsymbol{\bar{\nu}})^{T}\boldsymbol{\Sigma}^{-1}(\boldsymbol{\nu}-\boldsymbol{\bar{\nu}})}e^{-\frac{\left(\nu_{n}-a_{n}\right)^{2}}{2d_{n}}}\frac{\left(\nu_{n}-a_{n}\right)^{2}}{2d_{n}^{2}}\\
	& = & \frac{\langle\left(\nu_{n}-a_{n}\right)^{2}\rangle_{Q}}{2d_{n}^{2}}
\end{eqnarray*}
Setting the derivatives in (\ref{eq:DKLderiv}) to $0$ and assuming
$d_{n}\neq0$ we finally get

\begin{equation}
\begin{cases}
0 & =\frac{-\langle\nu_{n}\rangle_{Q^{\left(n\right)}}+a_{n}}{d_{n}}+\frac{\left\langle \nu_{n}\right\rangle _{Q}-a_{n}}{d_{n}}\\
0 & =-\frac{\langle\left(\nu_{n}-a_{n}\right){}^{2}\rangle_{Q^{\left(n\right)}}}{2d_{n}^{2}}+\frac{\langle\left(\nu_{n}-a_{n}\right)^{2}\rangle_{Q}}{2d_{n}^{2}}
\end{cases}
\end{equation}

\begin{equation}
\begin{cases}
\langle\nu_{n}\rangle_{Q^{\left(n\right)}} & =\langle\nu_{n}\rangle_{Q}\\
\langle\nu_{n}^{2}\rangle_{Q^{\left(n\right)}} & =\langle\nu_{n}^{2}\rangle_{Q}
\end{cases}
\end{equation}
and thus the moment matching condition in Eq. (13) turns out to be
equivalent to the KL-divergence minimization condition.

\subsection*{Supplementary note 2. Moments of the tilted distribution}

Let us compute $\left\langle \nu_{n}\right\rangle _{Q^{\left(n\right)}}$
and $\left\langle \nu_{n}^{2}\right\rangle _{Q^{\left(n\right)}}$
as

\begin{align}
\left\langle \nu_{n}\right\rangle _{Q^{\left(n\right)}} & =\frac{1}{\tilde{Z}_{Q^{\left(n\right)}}}\int d\nu_{n}\nu_{n}\psi_{n}(\nu_{n})\int\prod_{m\neq n}d\nu_{m}e^{-\frac{1}{2}(\boldsymbol{\nu}-\boldsymbol{\bar{\nu}})^{T}\boldsymbol{\Sigma}^{-1}(\boldsymbol{\nu}-\boldsymbol{\bar{\nu}})}\\
\left\langle \nu_{n}^{2}\right\rangle _{Q^{\left(n\right)}} & =\frac{1}{\tilde{Z}_{Q^{\left(n\right)}}}\int d\nu_{n}\nu_{n}^{2}\psi_{n}(\nu_{n})\int\prod_{m\neq n}d\nu_{m}e^{-\frac{1}{2}(\boldsymbol{\nu}-\boldsymbol{\bar{\nu}})^{T}\boldsymbol{\Sigma}^{-1}(\boldsymbol{\nu}-\boldsymbol{\bar{\nu}})}
\end{align}

Integrating out the multivariate Gaussian we obtain for the first
moment

\begin{equation}
\left\langle \nu_{n}\right\rangle _{Q^{\left(n\right)}}=\frac{1}{\tilde{Z}_{Q^{\left(n\right)}}}\int d\nu_{n}\nu_{n}q_{n}\left(\nu_{n}\right)
\end{equation}
where $q_{n}\left(\nu_{n}\right)$ is the marginal probability function

\begin{eqnarray}
q_{n}\left(\nu_{n}\right) & \propto & \psi_{n}(\nu_{n})e^{-\frac{(\nu_{n}-\bar{\nu}_{n})^{2}}{2\Sigma_{nn}}}\label{eq:marginal}\\
& \propto & \begin{cases}
e^{-\frac{\left(\nu_{n}-\bar{\nu}_{n}\right)^{2}}{2\Sigma_{nn}}} & \qquad\mathrm{if}\:\nu_{n}\in\left[v_{n}^{inf},\nu_{n}^{sup}\right]\\
0 & \mathrm{\qquad otherwise}
\end{cases}
\end{eqnarray}
Let us rewrite the non-zero part of (\ref{eq:marginal}) in standard
notation:

\begin{equation}
q_{n}\left(\nu_{n}\right)=\frac{\frac{1}{\Sigma_{nn}}\mathcal{{N}}\left(\frac{\nu-\bar{\nu}_{n}}{\sqrt{\Sigma_{nn}}}\right)}{\Phi\left(\frac{\nu_{n}^{sup}-\bar{\nu}_{n}}{\sqrt{\Sigma_{nn}}}\right)-\Phi\left(\frac{\nu_{n}^{inf}-\bar{\nu}_{n}}{\sqrt{\Sigma_{nn}}}\right)}\label{eq:gausstrunc}
\end{equation}
where the $\mathcal{{N}}\left(x\right)=\frac{1}{\sqrt{2\pi}}e^{-\frac{x^{2}}{2}}$
is the probability density function of the standard normal distribution
and $\Phi(x)=\int_{-\infty}^{x}\frac{e^{-\frac{y^{2}}{2}}}{\sqrt{2\pi}}dy=\frac{1}{2}\left[1+\erf\left(\frac{x}{\sqrt{2}}\right)\right]$
is its cumulative. In the following we will need the value of the
first two moments of this distribution that are given by:

\begin{eqnarray}
\langle\nu_{n}\rangle_{Q^{(n)}} & = & \bar{\nu}_{n}+\frac{\mathcal{{N}}\left(\frac{\nu_{n}^{inf}-\bar{\nu}_{n}}{\sqrt{\Sigma_{nn}}}\right)-\mathcal{{N}}\left(\frac{\nu_{n}^{sup}-\bar{\nu}_{n}}{\sqrt{\Sigma_{nn}}}\right)}{\Phi\left(\frac{\nu_{n}^{sup}-\bar{\nu}_{n}}{\sqrt{\Sigma_{nn}}}\right)-\Phi\left(\frac{\nu_{n}^{inf}-\bar{\nu}_{n}}{\sqrt{\Sigma_{nn}}}\right)}\sqrt{\Sigma_{nn}}\label{eq:meanQ(n)}\\
\langle\nu_{n}^{2}\rangle_{Q^{(n)}}-\langle\nu_{n}\rangle_{Q^{(n)}}^{2} & = & \Sigma_{nn}\left[1+\frac{\frac{\nu_{n}^{inf}-\bar{\nu}_{n}}{\Sigma_{nn}}\mathcal{{N}}\left(\frac{\nu_{n}^{inf}-\bar{\nu}_{n}}{\sqrt{\Sigma_{nn}}}\right)-\frac{\nu_{n}^{sup}-\bar{\nu}_{n}}{\sqrt{\Sigma_{nn}}}\mathcal{{N}}\left(\frac{\nu_{n}^{sup}-\bar{\nu}_{n}}{\sqrt{\Sigma_{nn}}}\right)}{\Phi\left(\frac{\nu_{n}^{sup}-\bar{\nu}_{n}}{\sqrt{\Sigma_{nn}}}\right)-\Phi\left(\frac{\nu_{n}^{inf}-\bar{\nu}_{n}}{\sqrt{\Sigma_{nn}}}\right)}\right.+\label{eq:varQ(n)}\\
&  & \left.-\left(\frac{\mathcal{{N}}\left(\frac{\nu_{n}^{inf}-\bar{\nu}_{n}}{\sqrt{\Sigma_{nn}}}\right)-\mathcal{{N}}\left(\frac{\nu_{n}^{sup}-\bar{\nu}_{n}}{\sqrt{\Sigma_{nn}}}\right)}{\Phi\left(\frac{\nu_{n}^{sup}-\bar{\nu}_{n}}{\sqrt{\Sigma_{nn}}}\right)-\Phi\left(\frac{\nu_{n}^{inf}-\bar{\nu}_{n}}{\sqrt{\Sigma_{nn}}}\right)}\right)^{2}\right]
\end{eqnarray}

Unfortunately when $\Sigma_{nn}\rightarrow0$ and thus $x\rightarrow+\infty$
several numeric issues occur when we compute (\ref{eq:meanQ(n)}),(\ref{eq:varQ(n)}).
We propose an expansion up to the $5^{th}$ order of these equations
to overcome this problem (see details in Supplementary note 5).

\subsection*{Supplementary note 3. Moments of $Q\left(\boldsymbol{\nu}|\boldsymbol{b}\right)$
	\label{sec:Moments-Q}}

Let us compute $\left\langle \nu_{n}\right\rangle _{Q}$ and $\left\langle \nu_{n}^{2}\right\rangle _{Q}$
as

\begin{eqnarray}
\left\langle \nu_{n}\right\rangle _{Q} & = & \frac{1}{\tilde{Z}_{Q}}\int d\nu_{n}\nu_{n}\phi_{n}(\nu_{n})\int\prod_{m\neq n}d\nu_{m}e^{-\frac{1}{2}(\boldsymbol{\nu}-\boldsymbol{\bar{\nu}})^{T}\boldsymbol{\Sigma}^{-1}(\boldsymbol{\nu}-\boldsymbol{\bar{\nu}})}\\
\left\langle \nu_{n}^{2}\right\rangle _{Q} & = & \frac{1}{\tilde{Z}_{Q}}\int d\nu_{n}\nu_{n}^{2}\phi_{n}(\nu_{n})\int\prod_{m\neq n}d\nu_{m}e^{-\frac{1}{2}(\boldsymbol{\nu}-\boldsymbol{\bar{\nu}})^{T}\boldsymbol{\Sigma}^{-1}(\boldsymbol{\nu}-\boldsymbol{\bar{\nu}})}
\end{eqnarray}

Integrating out the multivariate Gaussian we obtain 

\begin{eqnarray}
\left\langle \nu_{n}\right\rangle _{Q} & = & \frac{1}{\tilde{Z}_{Q}}\int d\nu_{n}\nu_{n}q_{n}\left(\nu_{n}\right)\\
\left\langle \nu_{n}^{2}\right\rangle _{Q} & = & \frac{1}{\tilde{Z}_{Q}}\int d\nu_{n}\nu_{n}^{2}q_{n}\left(\nu_{n}\right)
\end{eqnarray}
where $q_{n}\left(\nu_{n}\right)$ is the marginal probability function

\begin{eqnarray}
q_{n}\left(\nu_{n}\right) & \propto & e^{-\frac{(\nu_{n}-a_{n})^{2}}{2d_{n}}}e^{-\frac{(\nu_{n}-\bar{\nu}_{n})^{2}}{2\Sigma_{nn}}}\label{eq:marginalprob-1}
\end{eqnarray}
The proportional sign denotes that the normalization constant is missing.
Remembering that the product of two Gaussian distributions satisfy

\[
\mathcal{{N}}\left(x|\mu_{1},\sigma_{1}\right)\mathcal{{N}}\left(x|\mu_{2},\sigma_{2}\right)=\mathcal{{N}}\left(x|\mu,\sigma\right)
\]
where

\[
\begin{cases}
\frac{\mu}{\sigma} & =\frac{\mu_{1}}{\sigma_{1}}+\frac{\mu_{2}}{\sigma_{2}}\\
\frac{1}{\sigma} & =\frac{1}{\sigma_{1}}+\frac{1}{\sigma_{2}}
\end{cases}
\]
In our case, we obtain the following result for the first and second
(connected) moment of (\ref{eq:marginalprob-1}):

\begin{equation}
\begin{cases}
\langle\nu_{n}^{2}\rangle_{Q}-\langle\nu_{n}\rangle_{Q}^{2} & =\frac{1}{\frac{1}{d_{n}}+\frac{1}{\Sigma_{nn}}}\\
\langle\nu_{n}\rangle_{Q} & =\left(\frac{1}{d_{n}}+\frac{1}{\Sigma_{nn}}\right)^{-1}\left(\frac{a_{n}}{d_{n}}+\frac{\bar{\nu}_{n}}{\Sigma_{nn}}\right)
\end{cases}
\end{equation}

\subsection*{Supplementary note 4. Fast computation of $\boldsymbol{\Sigma}$
	and $\bar{\boldsymbol{\nu}}$ \label{sec:Fast-computation}}

Each time we update the parameters of one $\phi_{n}$ we need to build
a new matrix $\mathbf{D}$ and solve the system of equations in Eq.
(16) which requires the inversion of a big matrix of dimension $N\times N$.
Globally we need to invert $N$ times a large matrix per iteration
which severely affects the computational time. Here we present a scheme
by which we can invert one large matrix per iteration.

Let us define $\mathbf{D^{'}}$a diagonal matrix of elements $D_{nn}^{'}=\frac{1}{d_{n}}$
and $\boldsymbol{\Sigma}^{'}$, $\boldsymbol{\bar{\nu}}^{'}$the solutions
of

\begin{equation}
\begin{cases}
\boldsymbol{\Sigma}^{'-1} & =\beta\mathbf{S}^{T}\mathbf{S}+\mathbf{D}^{'}\\
\boldsymbol{\bar{\nu}}^{'} & =\boldsymbol{\Sigma}^{'}\left(\beta\mathbf{S}^{T}\boldsymbol{b}+\mathbf{D}^{'}\boldsymbol{a}\right)
\end{cases}\label{eq:}
\end{equation}

We aim at determining the values of $\boldsymbol{\Sigma}$ and $\bar{\boldsymbol{\nu}}$
entering in the computation of \textbf{$a_{n}$} and $d_{n}$ as functions
of $\boldsymbol{\Sigma}^{'}$and $\bar{\boldsymbol{\nu}}^{'}$that
can be computed per each iteration. Let us write for each flux $n$
the respective $\mathbf{D}$ matrix as $\mathbf{D=}\mathbf{D}'-\frac{1}{d_{n}}\boldsymbol{e}_{n}\boldsymbol{e}_{n}^{T}$
that must satisfy

\begin{equation}
\begin{cases}
\left(\beta\mathbf{S}^{T}\mathbf{S}+\mathbf{D}\right)\boldsymbol{\bar{\nu}} & =\beta\mathbf{S}^{T}\boldsymbol{b}+\mathbf{D}\boldsymbol{a}\\
\left(\beta\mathbf{S}^{T}\mathbf{S}+\mathbf{D^{'}}\right)\bar{\boldsymbol{\nu}}^{'} & =\beta\mathbf{S}^{T}\boldsymbol{b}+\mathbf{D^{'}}\boldsymbol{a}
\end{cases}\label{eq:D'sys}
\end{equation}
Take the first equation in (\ref{eq:D'sys}) and subtract to the second:

\begin{eqnarray}
\left(\beta\mathbf{S}^{T}\mathbf{S}+\mathbf{D^{'}}\right)\left(\boldsymbol{\bar{\nu}}-\boldsymbol{\bar{\nu}}^{'}\right)-\frac{1}{d_{n}}\boldsymbol{e}_{n}\boldsymbol{e}_{n}^{T}\boldsymbol{\bar{\nu}} & = & -\frac{1}{d_{n}}\boldsymbol{e}_{n}\boldsymbol{e}_{n}^{T}\boldsymbol{a}\\
\left(\beta\mathbf{S}^{T}\mathbf{S}+\mathbf{D^{'}}\right)^{-1}\left(-\frac{1}{d_{n}}a_{n}\boldsymbol{e}_{n}+\frac{1}{d_{n}}\bar{\nu}_{n}\boldsymbol{e}_{n}\right)+\bar{\boldsymbol{\nu}}^{'} & = & \boldsymbol{\bar{\nu}}
\end{eqnarray}
where it is possible to extract the $\bar{\nu}_{n}$ component as

\begin{eqnarray}
\bar{\nu}_{n}\left[1-\left(\beta\mathbf{S}^{T}\mathbf{S}+\mathbf{D^{'}}\right)_{nn}^{-1}\frac{1}{d_{n}}\right] & = & -D_{nn}^{'}a_{n}\left(\beta\mathbf{S}^{T}\mathbf{S}+\mathbf{D^{'}}\right)_{nn}^{-1}+\bar{\nu}_{n}^{'}\\
\bar{\nu}_{n} & = & \frac{-\frac{1}{d_{n}}a_{n}\left(\beta\mathbf{S}^{T}\mathbf{S}+\mathbf{D^{'}}\right)_{nn}^{-1}+\bar{\nu}_{n}^{'}}{1-\left(\beta\mathbf{S}^{T}\mathbf{S}+\mathbf{D^{'}}\right)_{nn}^{-1}\frac{1}{d_{n}}}\label{eq:mu_n}
\end{eqnarray}

Equivalently the diagonal elements of $\boldsymbol{\Sigma}$ satisfying
$\boldsymbol{\Sigma}^{-1}=\beta\mathbf{S}^{T}\mathbf{S}+\mathbf{D}$
can be computed as follows. We define\textbf{ $\boldsymbol{x}$} the
solution of equation$\boldsymbol{\Sigma}^{-1}\boldsymbol{x}=\boldsymbol{e}_{n}$
such that $\boldsymbol{x}$ is the $n^{th}$ column of $\boldsymbol{\Sigma}$;
thus $x_{n}=\Sigma_{nn}$. Now consider the homogeneous equation $\boldsymbol{\Sigma}^{'-1}\boldsymbol{x}^{'}=\boldsymbol{0}$
for $\boldsymbol{\Sigma}^{'-1}=\beta\mathbf{S}^{t}\mathbf{S}+\mathbf{D^{'}}$which
surely has solution $\boldsymbol{x}^{'}\mathbf{=}\mathbf{0}$. We
write the system of equations:

\begin{equation}
\begin{cases}
\left(\beta\mathbf{S}^{T}\mathbf{S}+\mathbf{D}\right)\boldsymbol{x} & =\boldsymbol{e}_{n}\\
\left(\beta\mathbf{S}^{T}\mathbf{S}+\mathbf{D}^{'}\right)\boldsymbol{x}^{'} & =\boldsymbol{e}_{n}
\end{cases}
\end{equation}
and we proceed with the same argument as before. Take the first equation
and subtract the second
\begin{eqnarray}
\left(\beta\mathbf{S}^{T}\mathbf{S}+\mathbf{D}^{'}-\frac{1}{d_{n}}\boldsymbol{e}_{n}\boldsymbol{e}_{n}^{T}\right)\boldsymbol{x}-\left(\beta\mathbf{S}^{T}\mathbf{S}+\mathbf{D}^{'}\right)\boldsymbol{x}^{'} & = & \boldsymbol{e}_{n}\\
\left(\beta\mathbf{S}^{T}\mathbf{S}+\mathbf{D}^{'}\right)\left(\boldsymbol{x}-\boldsymbol{x}^{'}\right)-\frac{1}{d_{n}}x_{n}\boldsymbol{e}_{n} & = & \boldsymbol{e}_{n}\\
\left(\beta\mathbf{S}^{T}\mathbf{S}+\mathbf{D}^{'}\right)\boldsymbol{x}^{'}+\boldsymbol{e}_{n}+\frac{1}{d_{n}}x_{n}\boldsymbol{e}_{n} & = & \left(\beta\mathbf{S}^{T}\mathbf{S}+\mathbf{D}^{'}\right)\boldsymbol{x}\\
\boldsymbol{x}^{'}+\left(1+\frac{1}{d_{n}}x_{n}\right)\left(\beta\mathbf{S}^{T}\mathbf{S}+\mathbf{D}^{'}\right)^{-1}\boldsymbol{e}_{n} & = & \boldsymbol{x}
\end{eqnarray}

Since $\boldsymbol{x}^{'}=\boldsymbol{0}$, the $n^{th}$component
of $\boldsymbol{x}$ will be:

\begin{eqnarray}
x_{n} & = & \left(1+\frac{1}{d_{n}}x_{n}\right)\left(\beta\mathbf{S}^{T}\mathbf{S}+\mathbf{D}^{'}\right)_{nn}^{-1}\\
x_{n} & = & \frac{\left(\beta\mathbf{S}^{T}\mathbf{S}+\mathbf{D}^{'}\right)_{nn}^{-1}}{1-\left(\beta\mathbf{S}^{T}\mathbf{S}+\mathbf{D}^{'}\right)_{nn}^{-1}\frac{1}{d_{n}}}.
\end{eqnarray}

Finally
\begin{equation}
\begin{cases}
\Sigma_{nn} & =\frac{\Sigma_{nn}^{'}}{1-\Sigma_{nn}^{'}\frac{1}{d_{n}}}\\
\bar{\nu}_{n} & =\frac{-\frac{1}{d_{n}}a_{n}\Sigma_{nn}^{'}+\bar{\nu}_{n}^{'}}{1-\Sigma_{nn}^{'}\frac{1}{d_{n}}}
\end{cases}\label{eq:Sigma_mu_smart}
\end{equation}

Components of $\boldsymbol{\bar{\nu}}$ different from $\bar{\nu}_{n}$
and all non-diagonal entries of $\boldsymbol{\Sigma}$ can be computed
following the same strategy; we do not report their expression here
since the update rules of $a_{n}$ and $d_{n}$ in Eq. (18) only depends
on terms in (\ref{eq:Sigma_mu_smart}).

\subsection*{Supplementary note 5. Asymptotic expansion of the first two moments
	of the tilted distribution\label{sec:Expansion}}

As already remarked in Supplementary note 2, the computation of the
first two moments of the tilted distribution $Q^{(n)}$ defined in
Eq. (15) turns out to be numerically difficult to compute in particular
in cases when we need to evaluate integrals over the tails of the
distributions. To overcome such difficulties, we resorted to an asymptotic
expansion to the $5^{th}$ order which accounts for both accuracy
and numerical stability in all conditions analyzed in our tests. The
idea is to start by noting that up to the required precision, in the
limit $x\rightarrow\infty,$ $\Phi(x)\simeq\frac{1}{2}-{\cal N}(x)\left(\frac{1}{x}-\frac{1}{x^{3}}+\frac{3}{x^{5}}-o\left(\frac{1}{x^{7}}\right)\right)$
so that:

\begin{eqnarray}
\frac{\phi(x_{0})-\phi(x_{1})}{\Phi(x_{1})-\Phi(x_{0})} & = & \frac{e^{-\frac{x_{0}^{2}}{2}}-e^{-\frac{x_{1}^{2}}{2}}}{e^{-\frac{x_{0}^{2}}{2}}\left(\frac{1}{x_{0}}-\frac{1}{x_{0}^{3}}+\frac{3}{x_{0}^{5}}\right)-e^{-\frac{x_{1}^{2}}{2}}\left(\frac{1}{x_{1}}-\frac{1}{x_{1}^{3}}+\frac{3}{x_{0}^{5}}\right)}\nonumber \\
& = & \frac{x_{0}^{5}x_{1}^{5}\left(1-e^{\frac{x_{1}^{2}-x_{0}^{2}}{2}}\right)}{e^{\frac{x_{1}^{2}-x_{0}^{2}}{2}}x_{1}^{5}(-3+x_{0}^{2}+x_{0}^{4})+x_{0}^{5}(3-x_{1}^{2}+x_{1}^{4})}\label{eq:asydphi1}\\
& = & \begin{cases}
\frac{x_{0}^{5}}{3-x_{0}^{2}-x_{0}^{4}} & \mathrm{\,\,\,for\,\,\,}(x_{1}^{2}-x_{0}^{2})\rightarrow\infty\\
\frac{x_{1}^{5}}{3-x_{1}^{2}+x_{1}^{4}} & \,\,\,\mathrm{for\,\,\,}(x_{1}^{2}-x_{0}^{2})\rightarrow-\infty
\end{cases}\label{eq:eq:asydphi2}
\end{eqnarray}

\begin{eqnarray}
\frac{x_{0}\phi(x_{0})-x_{1}\phi(x_{1})}{\Phi(x_{1})-\Phi(x_{0})} & = & \frac{x_{0}e^{-\frac{x_{0}^{2}}{2}}-x_{1}e^{-\frac{x_{1}^{2}}{2}}}{e^{-\frac{x_{0}^{2}}{2}}\left(\frac{1}{x_{0}}-\frac{1}{x_{0}^{3}}+\frac{3}{x_{0}^{5}}\right)-e^{-\frac{x_{1}^{2}}{2}}\left(\frac{1}{x_{1}}-\frac{1}{x_{1}^{3}}+\frac{3}{x_{1}^{5}}\right)}\nonumber \\
& = & \frac{x_{0}^{5}x_{1}^{5}\left(x_{1}-x_{0}e^{\frac{x_{1}^{2}-x_{0}^{2}}{2}}\right)}{e^{\frac{x_{1}^{2}-x_{0}^{2}}{2}}x_{1}^{5}(-3+x_{0}^{2}-x_{0}^{4})+x_{0}^{5}(3-x_{1}^{2}+x_{1}^{4})}\label{eq:asydvar1}\\
& = & \begin{cases}
\frac{x_{0}^{6}}{3-x_{0}^{2}-x_{0}^{4}} & \mathrm{\,\,\,for\,\,\,}(x_{1}^{2}-x_{0}^{2})\rightarrow\infty\\
\frac{x_{1}^{6}}{3-x_{1}^{2}-x_{1}^{4}} & \mathrm{\,\,\,for\,\,\,}(x_{1}^{2}-x_{0}^{2})\rightarrow-\infty
\end{cases}\label{eq:asydvar2}
\end{eqnarray}

Taking in consideration the expressions above and defining $x_{0}=\frac{\nu_{n}^{inf}-\bar{\nu}_{n}}{\sqrt{\Sigma_{nn}}},$
$x_{1}=\frac{\nu_{n}^{sup}-\bar{\nu}_{n}}{\sqrt{\Sigma_{nn}}}$, $s=\mathrm{sign(x_{0}x_{1})}$,
$m=\min(|x_{0}|,|x_{1}|)$, $\Delta^{2}=\frac{x_{1}^{2}-x_{0}^{2}}{2}$
, and calling 
\[
\gamma=\gamma\left(x_{0},x_{1}\right)=\frac{x_{0}^{5}x_{1}^{5}}{e^{\frac{x_{1}^{2}-x_{0}^{2}}{2}}x_{1}^{5}(-3+x_{0}^{2}+x_{0}^{4})+x_{0}^{5}(3-x_{1}^{2}+x_{1}^{4})}
\]
we can finally approximate the two first moments of the tilted distribution
in the following manner:

\begin{eqnarray*}
	\langle\nu\rangle_{Q^{(n)}} & = & \bar{\nu}+\sqrt{\Sigma_{nn}}\cdot\begin{cases}
		\frac{\mathcal{{N}}\left(x_{0}\right)-\mathcal{{N}}\left(x_{1}\right)}{\Phi\left(x_{1}\right)-\Phi\left(x_{0}\right)} & \,\,\,\mathrm{for\,\,\,m\leq6\,\,or\,\,s=-1}\\
		\gamma\left(1-e^{\frac{x_{1}^{2}-x_{0}^{2}}{2}}\right) & \,\,\,\mathrm{for\,\,\,m\geq6,\,s=1,\,\Delta^{2}<40}\\
		\frac{x_{0}^{5}}{3-x_{0}^{2}-x_{0}^{4}} & \,\,\,\mathrm{for\,\,\,m\geq6,\,s=1,\,\Delta^{2}\geq40}
	\end{cases}\\
	\langle\nu^{2}\rangle_{Q^{(n)}}-\langle\nu\rangle_{Q^{(n)}}^{2} & = & \Sigma_{nn}\cdot\begin{cases}
		1+\frac{x_{0}\mathcal{{N}}\left(x_{0}\right)-x_{1}\mathcal{{N}}\left(x_{1}\right)}{\Phi\left(x_{1}\right)-\Phi\left(x_{0}\right)}-\left(\frac{\mathcal{{N}}\left(x_{0}\right)-\mathcal{{N}}\left(x_{1}\right)}{\Phi\left(x_{1}\right)-\Phi\left(x_{0}\right)}\right)^{2} & \text{for}\,\,\,m\leq6\,\,or\,\,s=-1\\
		1+\gamma\left(x_{1}-x_{0}e^{\frac{x_{1}^{2}-x_{0}^{2}}{2}}\right)-\gamma^{2}\left(1-e^{\frac{x_{1}^{2}-x_{0}^{2}}{2}}\right)^{2} & \text{for}\,\,\,m\geq6,\,s=1,\,\Delta^{2}<40\\
		1+\frac{x_{0}^{6}}{3-x_{0}^{2}-x_{0}^{4}}-\left(\frac{x_{0}^{5}}{3-x_{0}^{2}-x_{0}^{4}}\right)^{2} & \text{for}\,\,\,m\geq6,\,s=1,\,\Delta^{2}\geq40
	\end{cases}
\end{eqnarray*}
Note that the generating series for $\Phi$ is of alternate signs
and one can easily see that upon considering only the $3^{rd}$ order,
the variance might turn to be negative. To overcome this difficulty,
only terms of order $1,5,\dots,4n+1$ must be considered, and so the
next useful approximation is of order $9^{th}$.

\subsection*{Supplementary note 6. Weighted Hit-and-Run \label{sec:Weighted-Hit-and-Run}}

Hit-and-Run is a Monte Carlo method that aims at uniformly sample
the feasible configuration space of fluxes. To add a non-uniform prior
such as $g\left(\nu_{i};a_{i},d_{i}\right)$ in Eq. (19) , one should
resort to an importance sampling generalization of HR (see e.g. \cite{chen_general_1996}).
However, the determination of the parameters $a_{i}$ and $d_{i}$
must be done by multiple HR convergences in some sort of gradient
descent scheme (in a procedure similar to Boltzmann learning), which
is deemed to be too time consuming. 

A seemingly simpler alternative is to perform a re-weighting of the
configurations explored by the uniform sampling in a way that the
HR marginal of the flux fits the experimental data. Calling $\nu_{i}$
the experimentally known flux and defining the re-weighting function
as $g\left(\nu_{i};a_{i},d_{i}\right)$, our scope is to tune $a$
and $b$ to reproduce the experimental marginal. More formally $g\left(\nu_{i};a_{i},d_{i}\right)$
is the exponential function of the unknown Lagrange multipliers enforcing
the constraint on the fixed marginal as the one introduced in Eq.
(19). One way of determining the two parameters and of performing
the sampling can be the following. The empirical first and second
moments of the re-weighted distribution will read

\begin{align}
\left\langle \nu_{i}\right\rangle _{g} & \simeq\sum_{\alpha=1}^{A}\nu_{i,\alpha}\frac{g\left(\nu_{i,\alpha};a_{i},d_{i}\right)}{W}\label{eq:mean-var-HR-constr}\\
\left\langle \nu_{i}^{2}\right\rangle _{g} & \simeq\sum_{\alpha=1}^{A}\nu_{i,\alpha}^{2}\frac{g\left(\nu_{i,\alpha};a_{i},d_{i}\right)}{W}\nonumber 
\end{align}
where the index $\alpha$ runs over all sampled configurations and
$W=\sum_{\alpha}g\left(\nu_{i,\alpha};a_{i},d_{i}\right)$ is the
normalization constant. This is a $2\times2$ system that needs to
be solved for variables $a_{i},d_{i}$. 

However, we will show in the following that the approximation in Eq.
\ref{eq:mean-var-HR-constr} is normally too rough to expect good
results with a reasonable number of sample points in most cases. To
give an example of the reliability of this procedure let us fix two
marginals of our choice for the biomass of a population of Escherichia
Coli described by the modified \textit{iJR904} model introduced in
section ``Results'' of the main text. The distribution of the unconstrained
biomass flux in this network is roughly a Gaussian distribution with
standard deviation $2\cdot10^{-2}$ and mean $\mu=0.03$ as shown
in Supplementary figure 1 (a). We will attempt to constraint the system
to two different ``observed'' biomass fluxes, both with standard
deviation $10^{-2}$ and means $0.09$ and $0.2$ respectively. 

We start by computing a uniformly weighted sampling set with standard
HR. We performed this with three different sets sizes $T=\left\{ 2.56\cdot10^{7},\,1.02\cdot10^{8},\,4.10\cdot10^{8}\right\} $.
In each of the two cases, we also apply constrained EP to fix the
marginal distribution to the observed one, obtaining parameters $a_{i}^{EP\left(0.09\right)},d_{i}^{EP\left(0.09\right)}$
and $a_{i}^{EP\left(0.2\right)},d_{i}^{EP\left(0.2\right)}$. Now
we can perform the re-weighting of the configurations according to
the functions $g\left(\nu_{i};a_{i}^{EP\left(0.09\right)},d_{i}^{EP\left(0.09\right)}\right)$
and $g\left(\nu_{i};a_{i}^{EP\left(0.2\right)},d_{i}^{EP\left(0.2\right)}\right)$.
We show in Supplementary figure 1 the re-weighted marginals of the
biomass flux (blue, green and yellow bars) along with the Gaussian
distributions with mean $0.09$ (Supplementary figure 1 (b)) and the
one with mean $0.2$ (Supplementary figure 1 (c)) that we would like
to retrieve (red line). In Supplementary figure 1 (b) we can notice
that as we increase the number of sampled points, the re-weighted
marginal very slowly approaches the desired one; differently in Supplementary
figure 1 (c) HR estimate fails to retrieve the fixed profile, indicating
that a much larger set of uniform sampling points would be needed. 

The reason is that in the second case, being the mean in the unconstrained
case equal to $0.03$, for an exponentially overwhelming fraction
of the HR points the value of the flux $\nu_{i}$ is far from the
mean value of the experimental distribution (and thus the associated
weight $g\left(\nu_{i};a,b\right)$ is exponentially small). As a
consequence the number of sampling points needed to reasonably sample
the constrained distribution becomes extremely large. This is what
happens to the experimental growth rate described in the ``Results''
section that cannot be recovered by this method in a feasible time. 

\section*{Supplementary figures}

\begin{figure}[H]
	\includegraphics[width=0.27\paperwidth]{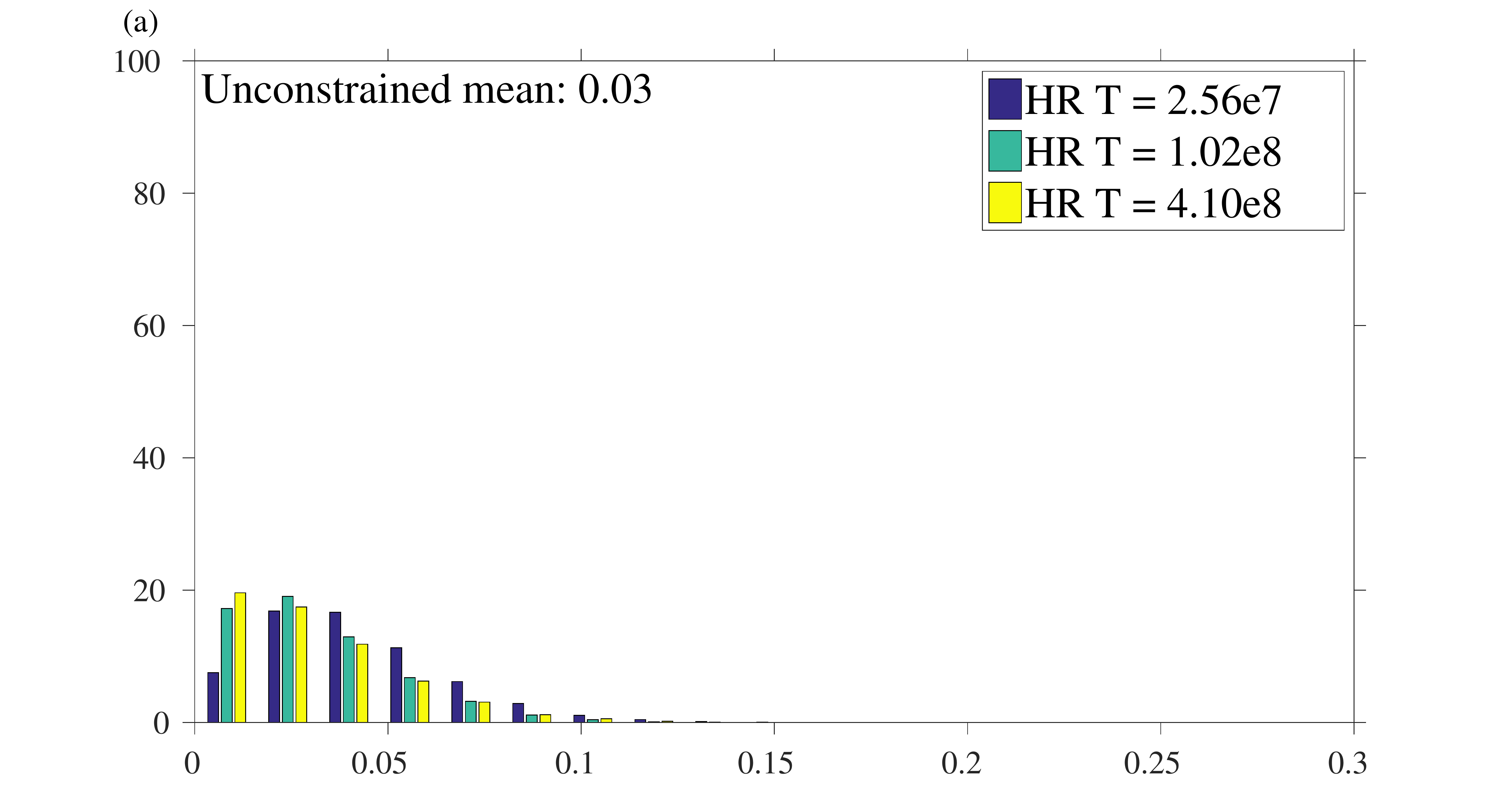}\includegraphics[width=0.33\textwidth]{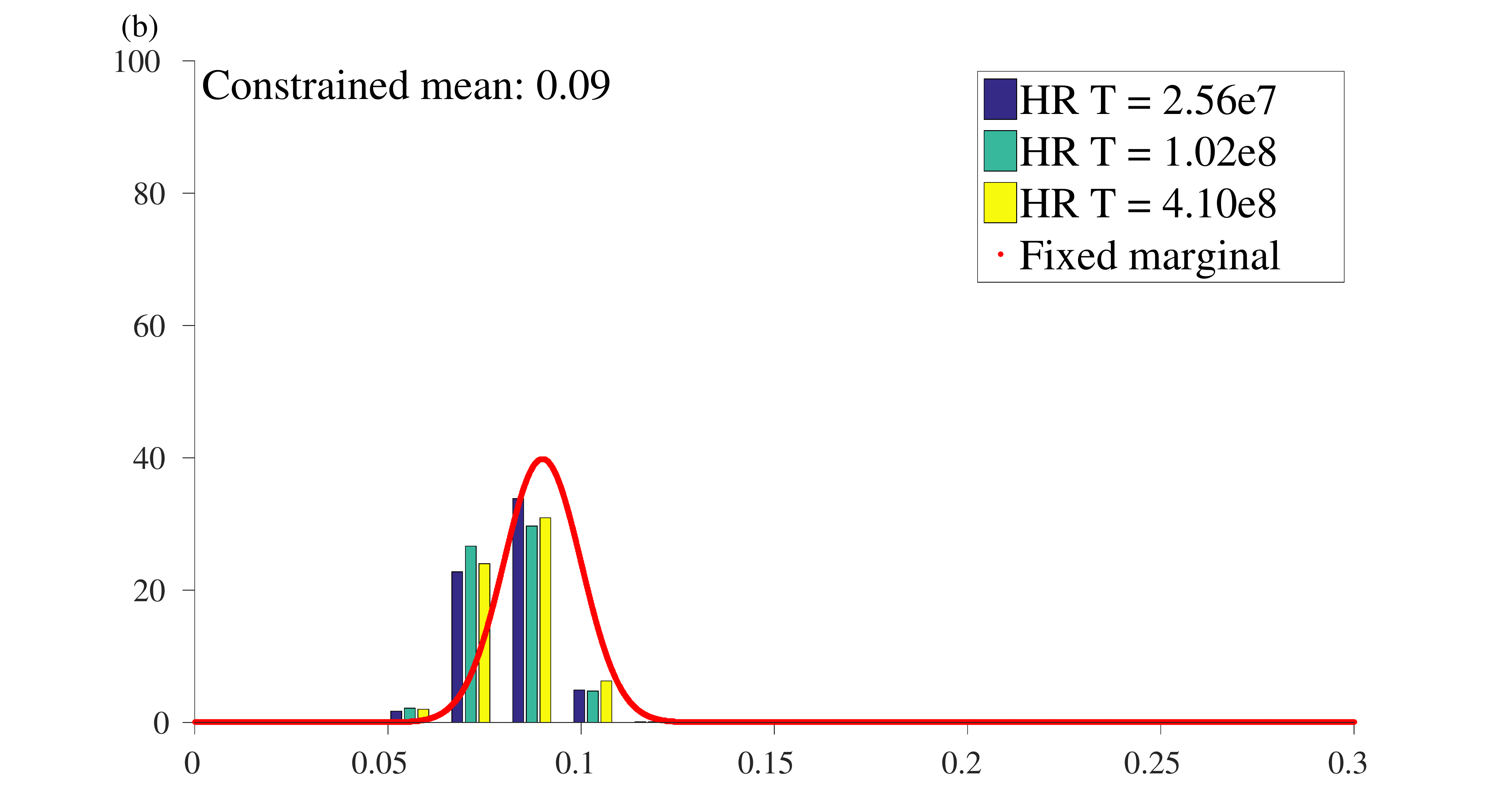}\includegraphics[width=0.33\textwidth]{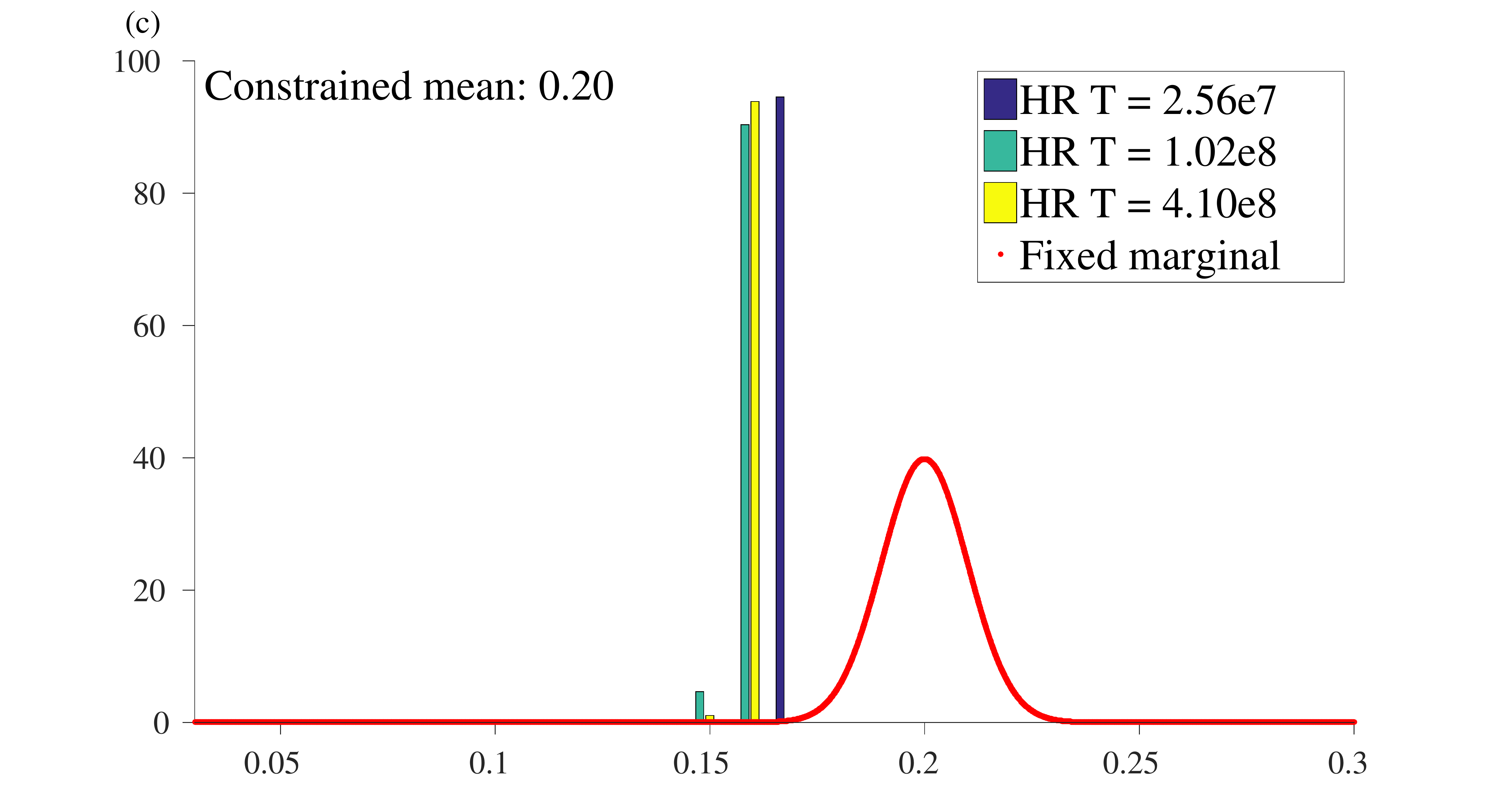}
	
	Supplementary figure 1: Marginal probability densities for the biomass
	flux computed through the re-weighting procedure (blue bars) and the
	fixed ones (red line) in two cases: (b) the fixed profile has mean
	0.09 while in (c) the mean has been shifted to 0.2. Fig. (a) shows
	the marginal in the unconstrained case.
	
	\vspace{5cm}
\end{figure}

\begin{figure}[H]
	\begin{centering}
		\includegraphics[width=0.6\textheight]{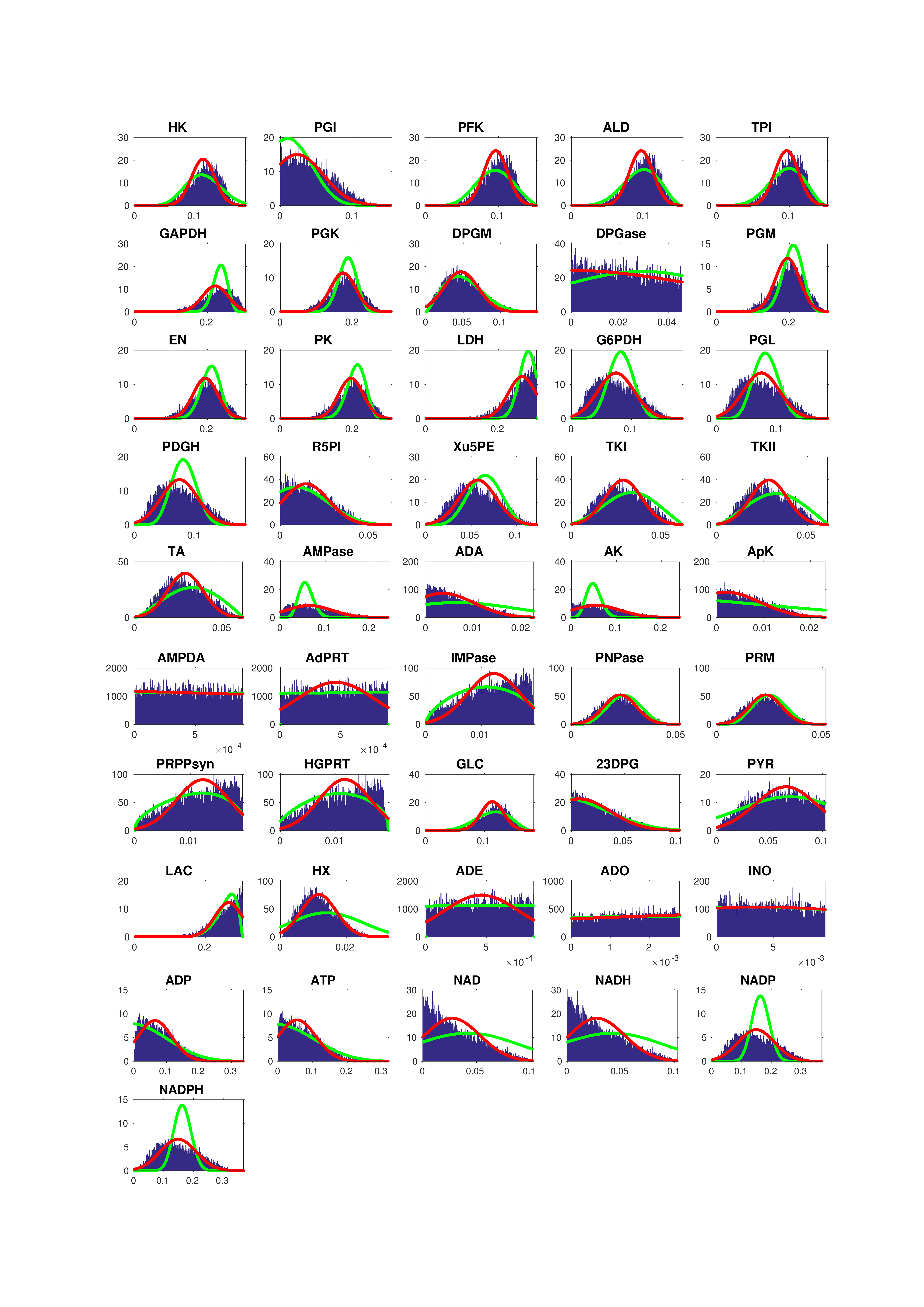}
		\par\end{centering}
	Supplementary figure 2: Marginals of the entire set of fluxes of red
	blood cell. The blue bars are obtained through HR sampling for $T\sim10^{8}$
	explored configurations; the green line is the prediction of the Belief
	propagation (BP) algorithm in \cite{fernandez-de-cossio-diaz_fast_2016}
	while the red line denotes the results of our EP algorithm. 
\end{figure}

\begin{figure}[H]
	\centering{}\includegraphics[width=1\textwidth]{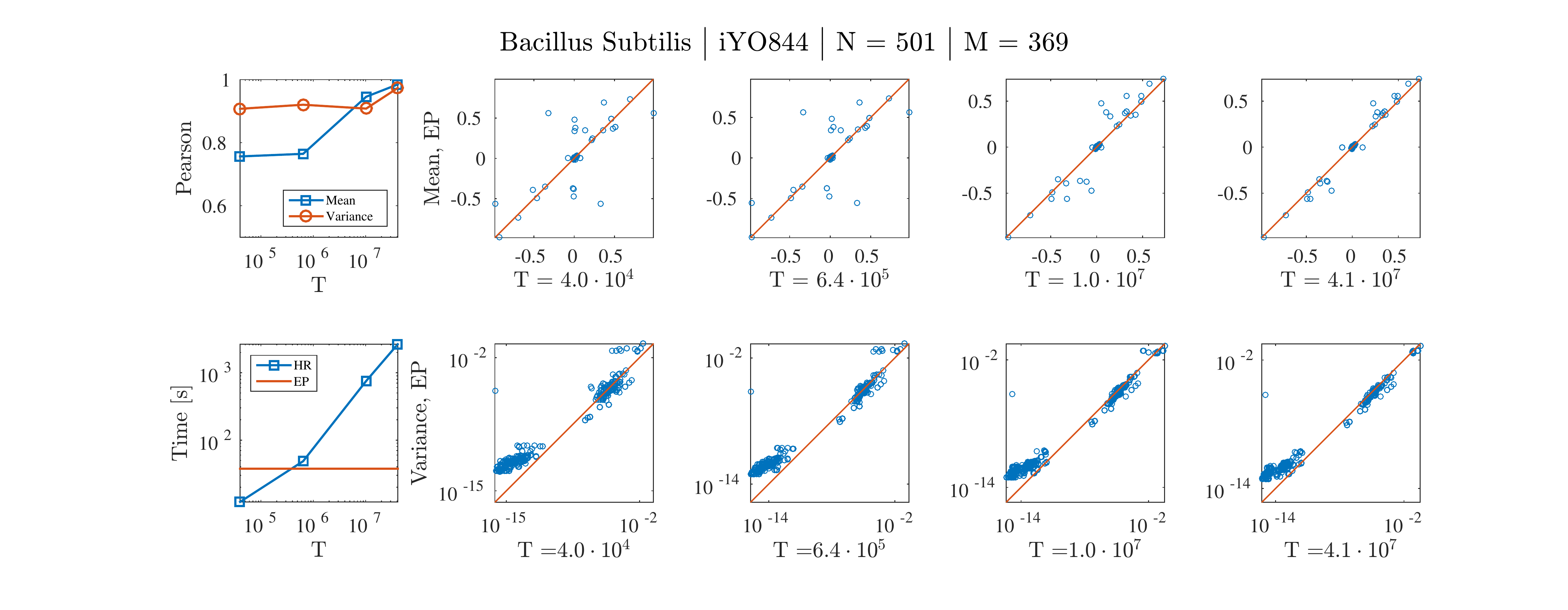}
\end{figure}

\begin{center}
	\begin{figure}[H]
		\centering{}\includegraphics[width=1\textwidth]{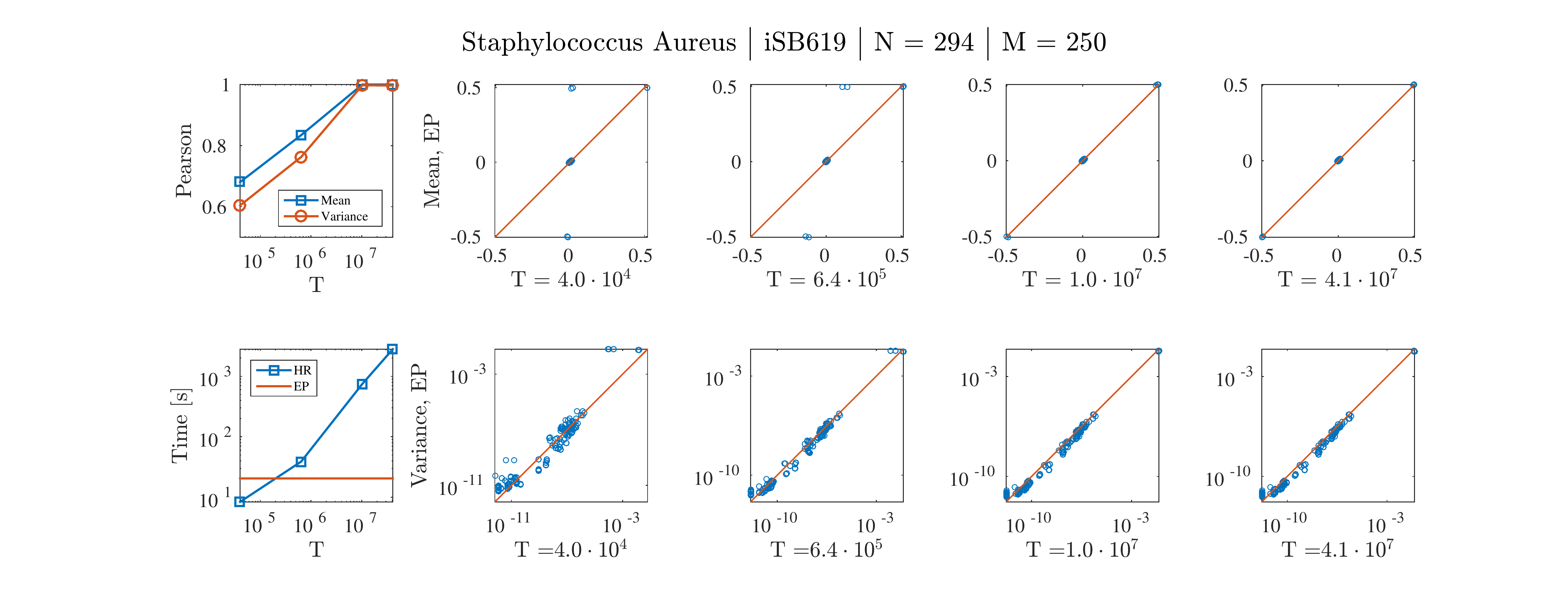}
	\end{figure}
	\par\end{center}

\begin{figure}[H]
	\centering{}\includegraphics[width=1\textwidth]{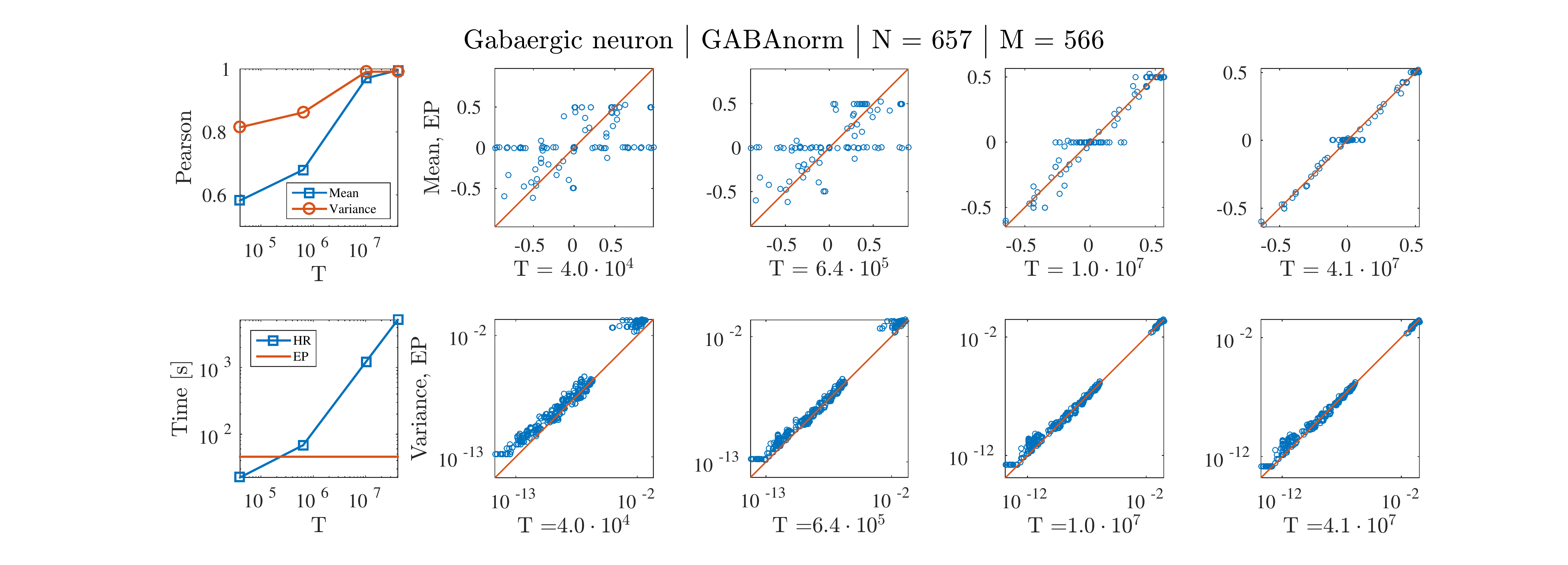}
\end{figure}

\begin{figure}[H]
	\centering{}\includegraphics[width=1\textwidth]{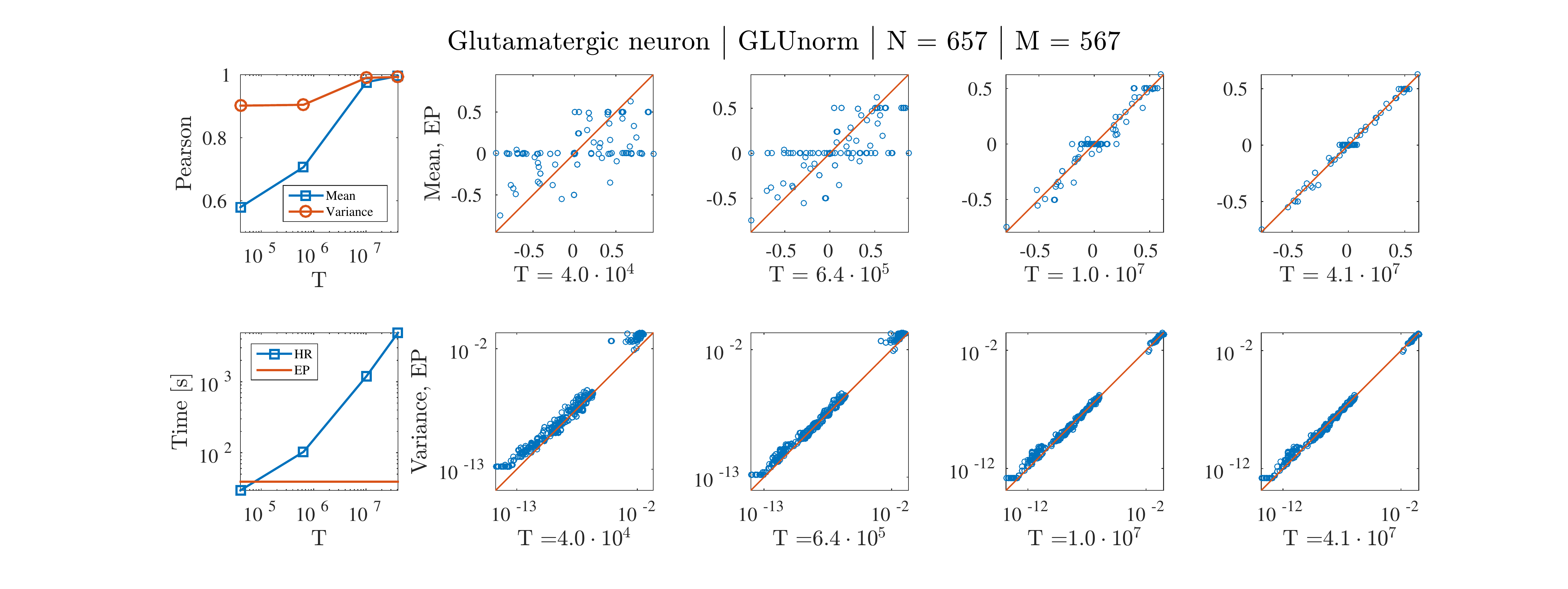}
\end{figure}

\begin{figure}[H]
	\centering{}\includegraphics[width=1\textwidth]{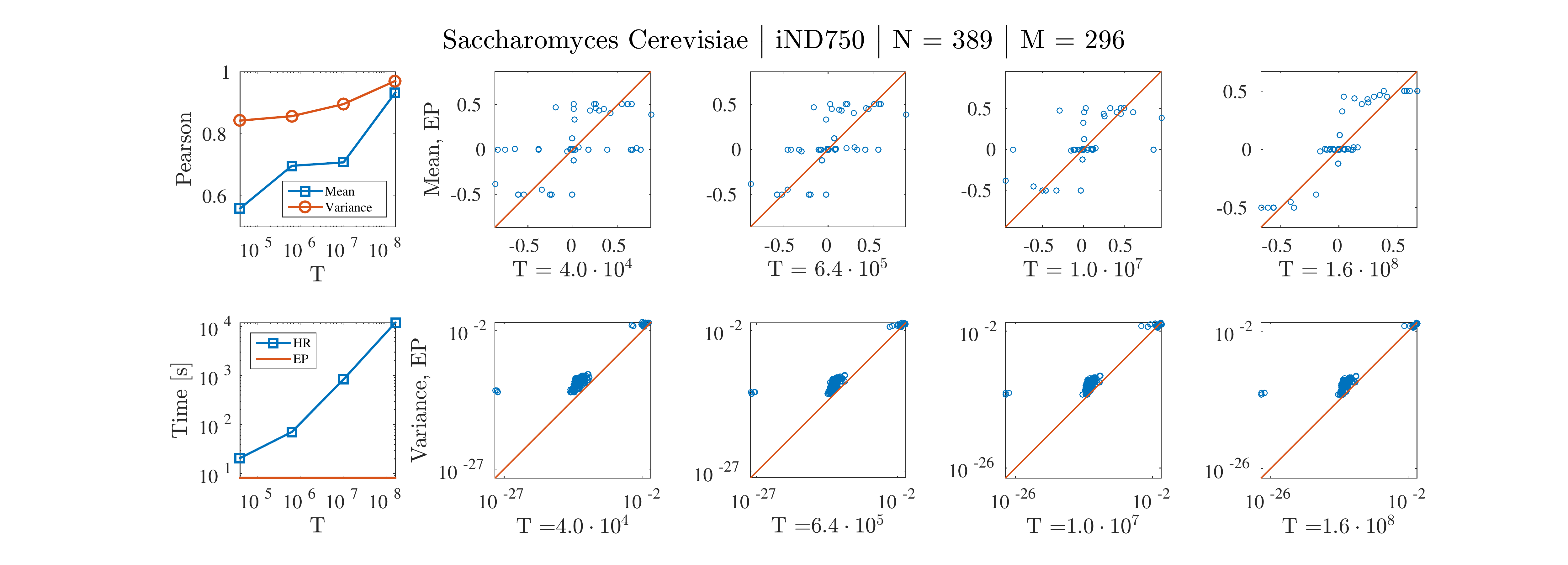}
\end{figure}

\begin{figure}[H]
	\centering{}\includegraphics[width=1\textwidth]{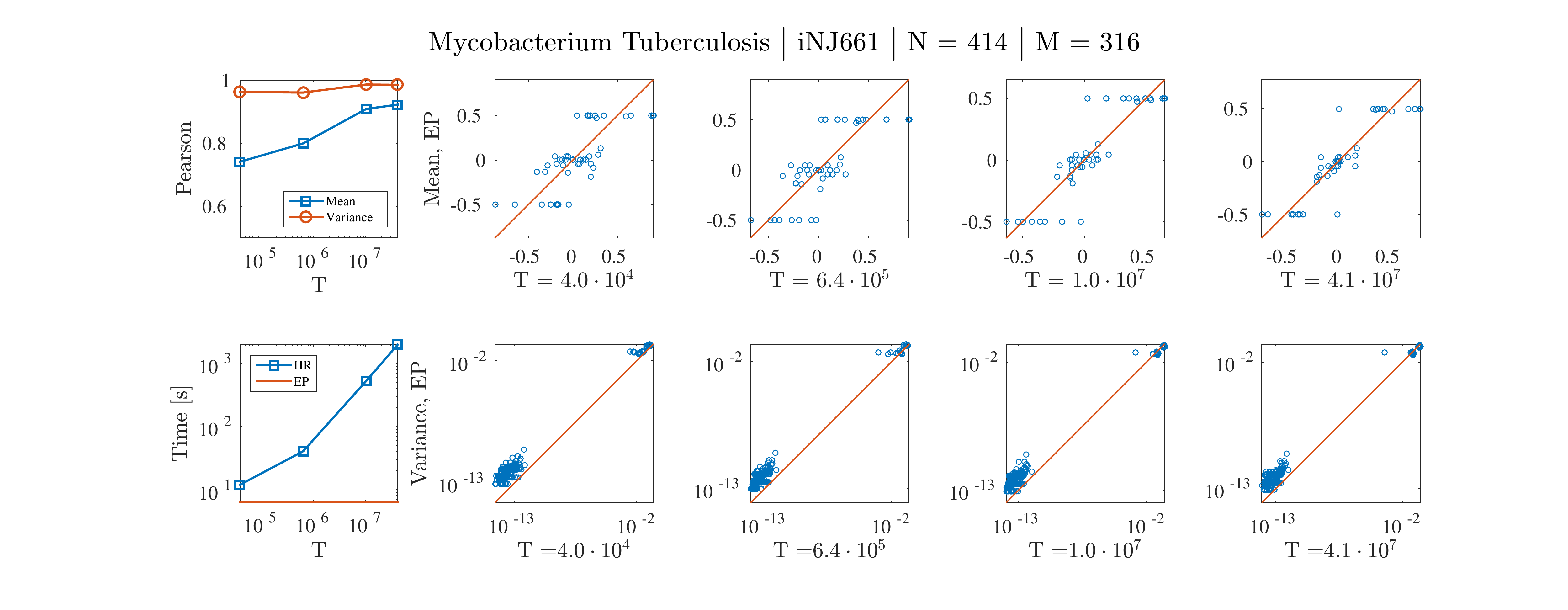}
\end{figure}

\begin{figure}[H]
	\centering{}\includegraphics[width=1\textwidth]{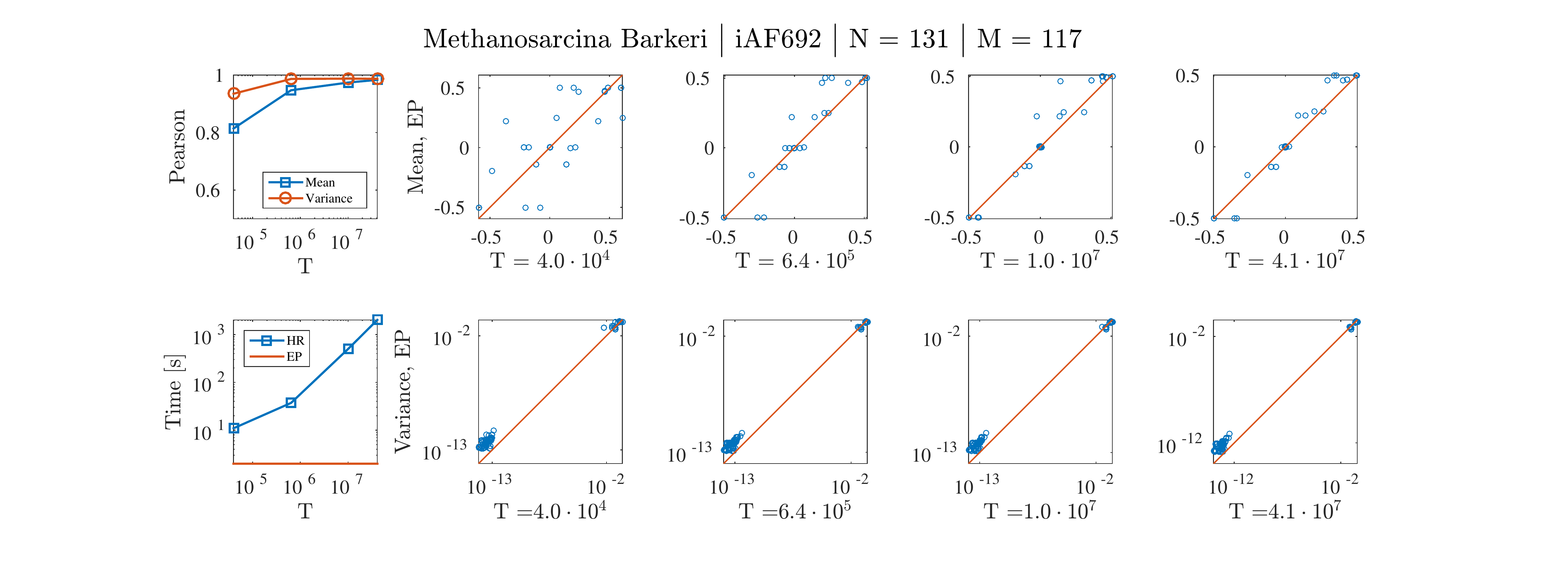}
\end{figure}

\begin{figure}[H]
	\centering{}\includegraphics[width=1\textwidth]{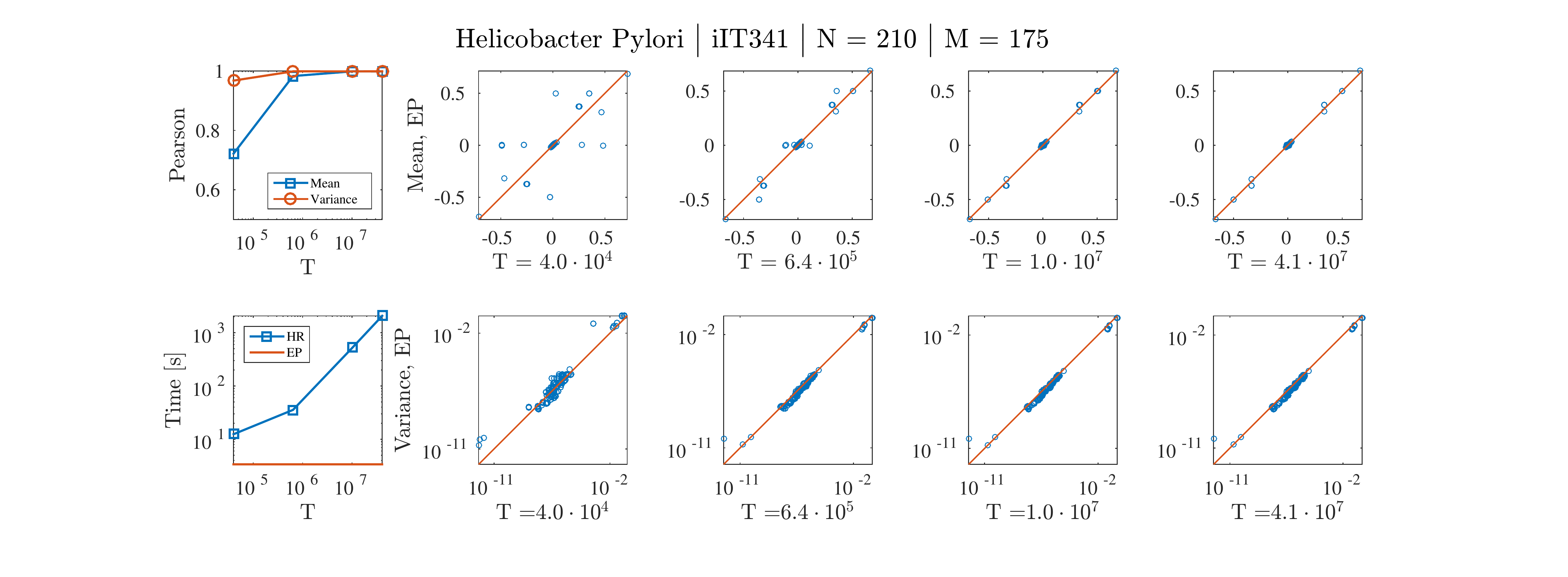}
\end{figure}

\begin{figure}[H]
	\centering{}\includegraphics[width=1\textwidth]{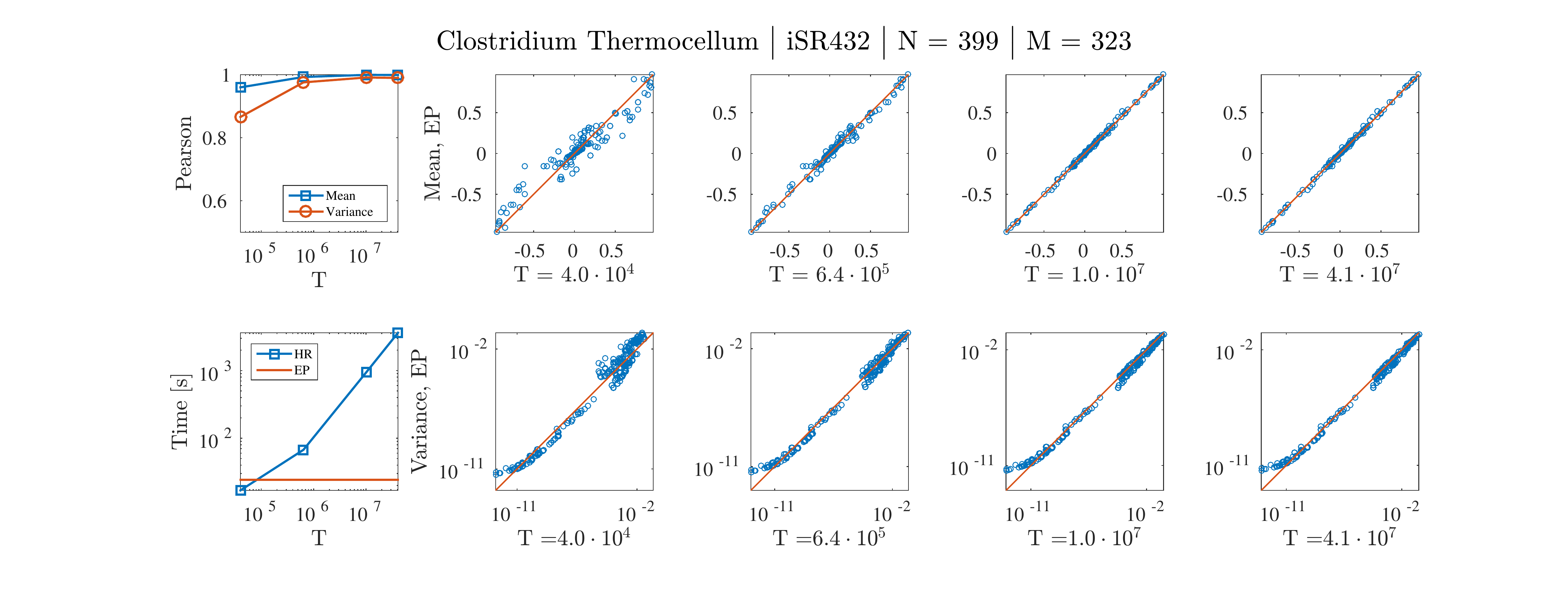}
\end{figure}

\begin{figure}[H]
	\begin{centering}
		\includegraphics[width=1\textwidth]{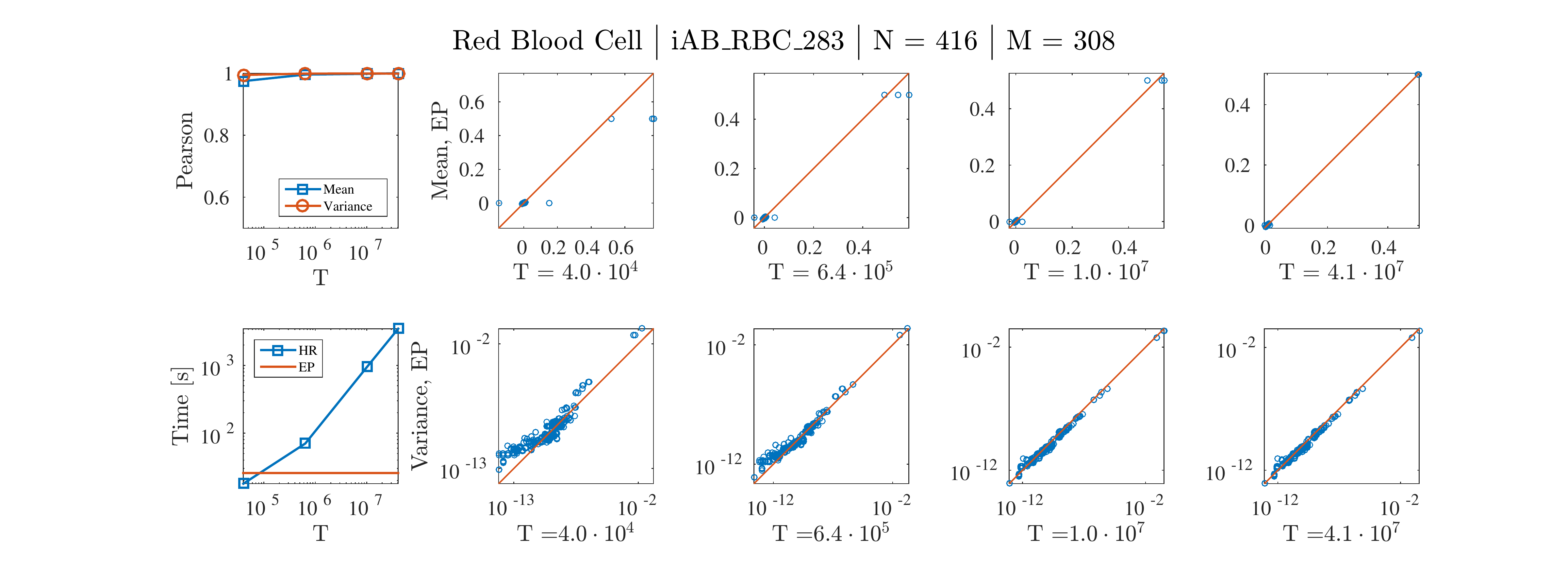}
		\par\end{centering}
	\centering{}Supplementary figure 3: Comparison between HR and EP for
	several models of large scale metabolic networks available in the
	Bigg Models database \cite{king2016bigg}. 
\end{figure}

\end{document}